\documentclass[a4paper, superscriptaddress, sd, amsfonts, amssymb, amsmath, preprint, showkeys, nofootinbib, nobibnotes]{revtex4-1}
\usepackage[english]{babel}
\usepackage[utf8]{inputenc}

\usepackage{xcolor}
\usepackage{graphicx}
\usepackage{dcolumn}
\usepackage{bm}
\usepackage{nicefrac}
\usepackage{units}
\usepackage{textcomp}
\usepackage[]{natbib} 
\usepackage{hyperref}
\hypersetup{colorlinks=true, citecolor=blue, urlcolor=cyan} 
\usepackage{cleveref}
\usepackage{verbatim}
\crefname{section}{§}{§§}
\Crefname{section}{§}{§§}
\usepackage{soul}
\usepackage[normalem]{ulem}

\begin{document}

\title{Structure and rheology of suspensions of spherical strain-hardening capsules}

\author{O. Aouane}
\email{o.aouane@fz-juelich.de}
\affiliation{Helmholtz Institute Erlangen-N\"{u}rnberg for Renewable Energy, Forschungszentrum J\"{u}lich, F\"{u}rther Stra{\ss}e 248, 90429 N\"{u}rnberg, Germany}

\author{A. Scagliarini}
\affiliation{Istituto per le Applicazioni del Calcolo 'M. Picone', IAC-CNR, Via dei Taurini 19, 00185 Roma, Italy} 
\affiliation{INFN, sezione Roma ``Tor Vergata'', via della Ricerca Scientifica 1, 00133 Rome, Italy}
\author{J. Harting}
\affiliation{Helmholtz Institute Erlangen-N\"{u}rnberg for Renewable Energy, Forschungszentrum J\"{u}lich, F\"{u}rther Stra{\ss}e 248, 90429 N\"{u}rnberg, Germany}
\affiliation{Department of Chemical and Biological Engineering and Department of Physics, Friedrich-Alexander-Universit\"at Erlangen-N\"urnberg, F\"{u}rther Stra{\ss}e 248, 90429 N\"{u}rnberg, Germany}

\date{\today} 

\begin{abstract}
We investigate the rheology of strain-hardening spherical capsules, from the dilute to the concentrated regime under a confined shear flow using three-dimensional numerical simulations.
We consider the effect of capillary number, volume fraction and membrane inextensibility on the particle deformation and on the effective suspension viscosity and normal stress differences of the suspension.
The suspension displays a shear-thinning behaviour which is a characteristic of soft particles such as emulsion droplets, vesicles, strain-softening capsules, and red blood cells.  
We find that the membrane inextensibility plays a significant role on the rheology and can almost suppress the shear-thinning. 
For concentrated suspensions a non-monotonic dependence of the normal stress differences on the membrane inextensibility is observed, reflecting a similar behaviour in the particle shape. The effective suspension viscosity, instead, grows and eventually saturates, for very large inextensibilities, approaching the solid particle limit. In essence, our results reveal that strain-hardening capsules share rheological features with both soft and solid particles depending on the ratio of the area dilatation to shear elastic modulus. Furthermore, the suspension viscosity exhibits a universal behaviour for the parameter space defined by the capillary number and the membrane inextensibility, when introducing the particle geometrical changes at the steady-state in the definition of the volume fraction.  
\end{abstract}

\maketitle

\section{\label{sec:introd}Introduction}
Capsules are closed elastic polymeric membranes, formed by cross-linking proteins to polysaccharides, and encapsulating a liquid droplet core \citep{levy1996coating,edwards1999serum}. Their typical diameter spans from a few nanometres to a few millimetres. Capsules are primarily used as controlled delivery systems of active substances with practical applications gearing around pharmaceutical industry \citep{donbrow1991microcapsules,de2010polymeric}, food processing \citep{sagis2008polymer}, cosmetics \citep{miyazawa2000preparation}, household products such as paints \citep{suryanarayana2008preparation}, while showing promising applications and future perspectives in areas such as thermal energy storage (\cite{sari2010preparation}), and injectable scaffolds for soft tissue regeneration \citep{munarin2010structural}. The release mechanisms of the active agents cover time scales going from a few seconds to days and can occur either via capsule burst or through slow and prolonged diffusion \citep{neubauer2014microcapsule}.
Capsules with a membrane made of polysiloxane will burst under continuous elongation \citep{walter2001shear,koleva2012deformation}, similar to droplets, while membranes formed with pure human serum albumin (HSA) or HSA-alginate can sustain very large deformations without rupture \citep{carin2003,de2015stretching} making them a good model for mimicking biological cells. The mechanical properties of the capsules can be probed by partial aspiration with a micropipette \citep{hochmuth2000micropipette}, atomic force microscopy \citep{fery2007mechanical}, compression tests \citep{chang1993experimentalextensional,rachik2006identification}, or in flow conditions by measuring the elongation of the particle and extracting the shear and area dilatation moduli using the appropriate hyperelastic constitutive law \citep{de2015stretching}. 

When subject to external stresses, capsules can exhibit a strain-softening or a strain-hardening behaviour depending on the composition and the fabrication protocol of their membrane. Those behaviours can be well recovered with hyperelastic constitutive laws such as the generalized Hooke, Mooney-Rivlin, Neo-Hookean, and Skalak laws \citep{de2015stretching,barthes2016motion}. The deformation and dynamics of a single capsule under different flow conditions have been thoroughly investigated by many authors: i) numerically  \citep{ramanujan1998deformation,navot1998elastic,lac2004spherical,li2008front,bagchi2009dynamics,dodson2009dynamics,farutin20143d,dupont2016stable,boedec2017isogeometric}, ii) analytically \citep{barthes1980motion,barthes1981time,vlahovska2011dynamics}, and iii) experimentally \citep{chang1993experimentalshear,chang1993experimentalextensional,de2015stretching,de2016tank,haener2020deformation}. Under a simple shear flow, initially non-spherical capsules can exhibit several complex dynamics including the steady and oscillating tank-treading, tumbling, and swinging motions (also called vacillating-breathing in the literature). These dynamics are the result of the interplay between different parameters like the capillary number, the confinement, and the viscosity contrast between the inner and the suspending fluids \citep{bagchi2009dynamics,walter2011ellipsoidal,vlahovska2011dynamics}. Conversely, a spherical capsule exhibits only a steady tank-treading motion characterised by a fixed orientation angle with respect to the flow and a steady-state deformed shape while the membrane undergoes a tank-treading motion \citep{bagchi2011dynamic,de2016tank}.

Most end-use applications of capsules involve many particle interactions, and often different coupled time scales, with non-linearity appearing already on the level of the single particle mechanics. This level of complexity requires a numerical approach to understand the behaviour of suspensions of capsules, the correlation between their microstructure and rheology, and how it compares with well studied systems such as solid particles and emulsions of drops.

The rheology of a dilute suspension of rigid spheres in an unbounded shear flow has been addressed analytically in the original work of \cite{einstein1906neue,einstein1911berichtigung} and extended to the second order by \cite{batchelor1972determination}
to include pair hydrodynamic interactions. Empirical, semi-empirical, and analytical models were proposed in the literature to predict the change of the relative viscosity with the volume fraction of rigid sphere suspensions from the dilute to the concentrated regimes \citep{eilers1941viskositat,mooney1951viscosity,maron1956application,krieger1959mechanism,frankel1967viscosity}.
Experiments and numerical simulations have revealed that a suspension of rigid particles exhibits shear-thinning, Newtonian, and shear-thickening behaviour, respectively, as the shear rate is increased. 
Both normal stress differences, $N_1$ and $N_2$, have been reported to bear a negative sign with a larger magnitude for $N_2$ with respect to $N_1$. 
The sign of $N_1$ is still subject to discussions because its magnitude is very small. 
Numerical simulations performed by \cite{sierou2002rheology} and \cite{gallier2014rheology,gallier2016effect} have pointed toward the possibility of two distinct physical origins of both normal stress differences. 
$N_2$ is associated to particle-particle collisions, while $N_1$ is mostly of hydrodynamic nature and is significantly affected by the presence or absence of boundaries. 
Detailed reviews on the rheology of solid particle suspensions can be found in \cite{stickel2005fluid}), \cite{mueller2010rheology}, and \cite{guazzelli2018rheology}.

Non-Newtonian behaviour in the form of shear-thinning has been reported for emulsions of drops, strain-softening capsules, and closed phospholipid bilayer membranes (vesicles). These three classes of deformable particles are characterized by a thin, continuous, and impermeable interface encapsulating an internal fluid. However, the interface mechanical properties are different. They all exhibit a negative $N_2$ and a positive $N_1$, and unlike rigid particles, the magnitude of $N_1$ is larger than $N_2$ which most probably indicates a more dominant role of hydrodynamic interactions as compared to particle-particle collisions \citep{loewenberg1996numerical,clausen2011rheology,matsunaga2016rheology,vlahovska2007dynamics,zhao2013dynamics}. 
In analogy to drops, a capillary number, quantifying the shear elastic resistance of the membrane to external stresses, has been widely used in the literature of capsules regardless of the nature of the hyperelastic law and the number of moduli characterizing the membrane mechanics. \cite{bagchi2010rheology} have shown that for a single strain-hardening capsule under a simple shear flow, the ratio of the area dilatation to shear elastic moduli, characterising the local inextensibility of the membrane, leads to some atypical effects on the intrinsic viscosity and the shear-thinning behaviour. 
For a non-dilute suspension of anisotropic strain-hardening capsules, \cite{gross2014rheology} have reported that for small capillary number the impact of the ratio of the area dilatation to shear elastic moduli on the suspension rheology is negligible.
 
This paper is devoted to study the dynamics of suspensions of model soft particles with a strain-hardening character, such as certain types of capsules \citep{carin2003}, focusing on the role played by a material property of the membrane, namely its inextensibility, on the capsule deformation
and on the suspension rheology, at varying the concentration of the dispersed phase and applied shear.
To the best of our knowledge the contribution of the local inextensibility of the membrane to the rheology of semi-dilute and concentrated suspensions of initially spherical strain-hardening capsules has so far not been studied.

The paper is organised as follows. We describe the numerical method in \S{2}
(benchmarks are provided in the Appendix).
We then present and discuss our numerical results for the dilute, semi-dilute and concentrated regimes as a function of the parameter $C$ quantifying the inextensibility of the membrane and the capillary number $Ca$ in \S{3}. The concluding remarks and further discussions are given in \S{4}. 
  
\section{\label{sec:method}Numerical method}

\subsection{Lattice Boltzmann Method (LBM)}

The Navier-Stokes equations are recovered in the limit of small Mach and Knudsen numbers by the
LBM, which is based on the discretisation of the Boltzmann-BGK \citep{bhatnagar1954model} equation in time and phase space (\cite{benzi1992lattice,succi2001lattice,kruger2017lattice}).
The LBM describes the evolution of the single particle distribution function $f_i(\mathbf{x},t)$ at a position $\mathbf{x}$ and time $t$ with a microscopic velocity $\mathbf{c}_i$, where $i = 1\dots Q$, on a regular $D$-dimensional lattice in discrete time steps $\Delta t$. We consider in this work a $D3Q19$ model corresponding to a three-dimensional lattice with $Q=19$ velocities. The lattice Boltzmann equation reads:
\begin{equation}
  f_i(\mathbf{x}+\mathbf{c}_i\Delta t, t + \Delta t) - f_i(\mathbf{x},t) = \Omega_i(\mathbf{x},t) + F_i(\mathbf{x},t) \Delta t,
  \label{eq:lbe}
\end{equation}
with 
\begin{equation}
   \Omega_i(\mathbf{x},t) = -\frac{\Delta t}{\tau}[f_i(\mathbf{x},t)-f_i^{eq}(\mathbf{x},t)],
\end{equation}
and
\begin{equation}
      f_i^{eq}(\mathbf{x},t) = \omega_i \rho \left[1 + \frac{(\mathbf{c}_i \cdot \mathbf{u})}{c_s^2} + \frac{1}{2}\frac{(\mathbf{c}_i \cdot \mathbf{u})^2}{c_s^4} - \frac{1}{2}\frac{|\mathbf{u}|^2}{c_s^2} \right], 
\end{equation}
where $\tau$ is a relaxation time related to the fluid dynamic viscosity by
$\mu_0 = \rho_0 c_s^2\left(\tau - \frac{\Delta t}{2}\right)$ ($\rho_0$ being the mean fluid density). $c_s=\frac{1}{\sqrt{3}} \frac{\Delta x}{\Delta t}$ denotes the lattice speed of sound, $\Delta x$ is the lattice constant, $\omega_i$ are lattice weights and $f_i^{eq}(\mathbf{x},t)$ is the equilibrium probability distribution function, depending on the fluid density $\rho$ and velocity $\mathbf{u}$ fields as a truncated expansion of the Maxwell-Boltzmann distribution 
(valid at small Mach number, $Ma = |\mathbf{u}|/c_s \ll 0.1$). 
$F_i$ in the RHS of equation \ref{eq:lbe} is a source term accounting for any external or internal force and will be used here to incorporate the forces exerted by the membrane on the fluid through the immersed boundary method.  
For a D3Q19 LBM, the lattice weights $\omega_i$ read as $1/3$, $1/18$ and $1/36$ for $i=1$, $i=2\dots7$, and $i=8\dots19$, respectively. 
The macroscopic fluid density $\rho$ and velocity $\mathbf{u}$ are deduced from the moments of the discrete probability distribution functions as 
\begin{equation}
   \rho = \sum_{i=1}^{19} f_i (\mathbf{x},t), \quad \text{and} \hspace{0.25cm}  \mathbf{u} = \sum_{i=1}^{19} f_i (\mathbf{x},t)\mathbf{c}_i / \rho.
\end{equation}
For convenience we set the lattice constant, the time step, the mean fluid density, and the relaxation time to unity.

\subsection{\label{subsec:membrane-model}Membrane model}

The capsule is modelled as a two-dimensional hyperelastic thin shell encapsulating an inner fluid and suspended in an outer fluid. The interior and exterior fluids are Newtonian with equal densities and viscosities. 
The surface of the capsule is discretized into a triangular mesh and endowed with a resistance to in-plane deformations. 
The membrane Lagrangian variables are defined on a moving curvilinear mesh with coordinates $(\xi_1,\xi_2)$, freely evolving on the Cartesian mesh on which lies the Eulerian fluid variables.  
We consider the case of a strain-hardening membrane using the hyperelastic law introduced by \cite{skalak1973modelling}, where the in-plane elastic deformations are governed by shear and area dilatation resistances. 
In terms of the deformation invariants, $I_1=\lambda_1^2 + \lambda_2^2 -2$ and $I_2 = \lambda_1^2 \lambda_2^2 -1$, where $\lambda_1$ and $\lambda_2$ are the principal stretching ratios on an element of the membrane surface, the strain energy over the surface of the capsule ($\Omega_S$) reads as
\begin{equation}
   E_s = \int_{\Omega_S} \frac{G_s}{4}[I_1^2 + 2 I_1 -2 I_2 + C I_2^2] d{\Omega_S}.
   \label{eq:local_strain}
\end{equation}

Here, $G_s$ is the elastic shear modulus and $C$ is a constant related to the strain-hardening character of the membrane through the scaled area dilatation modulus $G_a$ such as $G_a/G_s= 1 + 2C$. In other words, increasing the value of $C$ enhances the local inextensibility of the membrane (whence, we will refer to $C$, hereafter, as membrane inextensibility) and makes the capsule more strain-hardening.
In the small deformation limit, $C$ and the surface Poisson ratio ($\nu_s$) are related by $C = \frac{\nu_s}{1-\nu_s}$ (\cite{barthes2002effect}). 
The elastic deformations on the surface of the particle are evaluated numerically using a linear finite element method following the approach described in \cite{kruger2011efficient}.
In addition to shear elasticity and area dilatation, capsules may also exhibit a resistance to out-of-plane deformations (bending). The existence of a non-negligible bending energy depends on the protocol used to fabricate the capsule and the composition of the encapsulating membrane \citep{de2014mechanical,de2015stretching,gubspun2016characterization}. Although it can be of interest to investigate the interplay between the bending and the shear elasticity under different flow conditions, we choose to focus, here, on the role of the area dilatation, which has been somehow overlooked in the existing literature, and we consider, thus, capsules without bending resistance.
The volume of the capsules is prescribed using a penalty function that reads as  
\begin{equation}
   E_v = \frac{\kappa_v}{2}\frac{[V-V_0]^2}{V_0},
   \label{eq:volume_conservation}
\end{equation}
where $\kappa_v$ is a modulus that controls the deviation from the reference volume $V_0$ corresponding to the stress-free shape.
The membrane force is calculated on each membrane node ($\mathbf{X}^{m}$) using the principle of virtual work, such that 
\begin{equation}
    \mathbf{F}^{m} = -\frac{\partial E(\mathbf{X}^{m})}{\partial \mathbf{X}^{m}},
\end{equation}
where $E = E_s + E_v$ is the membrane total energy.

To avoid overlap between capsules in relatively dense systems, 
we introduce a short range repulsive force intended to mimic the normal component of the lubrication force. 
The repulsive force acts when the distance between two nodes from different capsules is below the cut-off distance $\delta_0$, which is set here to the minimum value of $1\Delta x$ corresponding to the fluid resolution limit of the LBM, and vanishes at a node-to-node distance $d_{ij} \ge \delta_0$. Its expression is given by
\begin{equation}
\mathbf{F}_{rep} = \left\{
\begin{array}{rcr}
&\bar{\epsilon}[(\frac{\Delta x}{d_{ij}})^2 - (\frac{\Delta x}{\delta_0})^2] \frac{\mathbf{d}_{ij}}{d_{ij}} \quad \text{if} \quad d_{ij} < \delta_0 \\
&\mathbf{0} \quad \text{if} \quad d_{ij} \ge \delta_0
\end{array}
\right. ,
\label{eq:frep}
\end{equation} 
where $\bar{\epsilon}$ is the interaction strength with the dimension of a force. 
Equation (\ref{eq:frep}) was suggested by \cite{glowinski2001fictitious} for suspension of rigid particles, and used by \cite{gross2014rheology} to study the rheology of very dense suspensions of red blood cells in a shear flow. Other contact models based for example on an exponential repulsive force or a Lennard-Jones potential can also be used \citep{guckenberger2016boundary,buxton2005newtonian,macmeccan2007mechanistic}. The effect of the short range repulsive force on the rheology of strain-hardening capsules is discussed in Subsection \ref{sec:appendix-repulsive_force} of the Appendix.

\subsection{Membrane dynamics}

For the fluid-membrane coupling, we use the immersed boundary method (IBM) \citep{peskin2002immersed}. In the IBM the Lagrangian massless 
nodes are interacting with the Eulerian fluid nodes using an interpolation function in a two-way coupling scheme: the Lagrangian membrane forces calculated on the curvilinear mesh are distributed to the surrounding Eulerian fluid nodes on the fixed Cartesian mesh by a smoothed approximation of the delta function, where they enter the discretized lattice Boltzmann equation (\ref{eq:lbe}) as an external force term. The new fluid velocities are obtained after solving the LBM equation (\ref{eq:lbe}). The capsules are advected with the fluid velocity, where the velocity of each Lagrangian node on the membrane is interpolated from the surrounding Eulerian fluid node velocity using the same scheme as for the spreading of the forces. The distribution of the membrane forces $\mathbf{F}^{m}$ located at position $\mathbf{X}^{m}(\xi_1,\xi_2,t)$ to the adjacent fluid nodes is given by
\begin{equation}
  \mathbf{f}(\mathbf{x},t) = \int_{\Omega_S} \mathbf{F}^{m}(\xi_1,\xi_2,t) \delta(\mathbf{x}-\mathbf{X}^{m}(\xi_1,\xi_2,t))d{\Omega_S},
  \label{eq:ibm_force_spreading}
\end{equation}
where $\delta$ is a three-dimensional approximation of the Dirac delta function, and $\mathbf{f}(\mathbf{x},t)$ is the force density acting on the fluid at the Eulerian node $\mathbf{x}(x_1,x_2,x_3)$. 
Equation (\ref{eq:ibm_force_spreading}) is incorporated into equation (\ref{eq:lbe}) in a similar manner to an external body force ({\it{e.g.}} gravity) as follows
\begin{equation}
   F_i(\mathbf{x},t) = \left(1 - \frac{1}{2\tau} \right) \omega_i \left(\frac{\mathbf{c}_i-\mathbf{u}}{c_s^2}+\frac{\mathbf{c}_i \cdot \mathbf{u}}{c_s^4}\mathbf{c}_i\right)\cdot\mathbf{f}(\mathbf{x},t).
\end{equation}
 
The velocity of the membrane is obtained from the local Eulerian fluid velocity as
\begin{equation}
    \frac{\partial \mathbf{X}^{m}}{\partial t} = \mathbf{u}(\mathbf{X}^m,t+\Delta t) = \int_{\Omega_D} \mathbf{u}(\mathbf{x},t+\Delta t) \delta(\mathbf{x}-\mathbf{X}^{m}(\xi_1,\xi_2,t)) dx^3,
    \label{eq:ibm_velocity_interpolation}
\end{equation}
where $\Omega_D$ represents the whole fluid domain. Equation (\ref{eq:ibm_velocity_interpolation}) enforces a no-slip boundary condition on the membrane, although in practice a volume drift is observed with time, and thus the need to use a penalty function on the volume (see equation (\ref{eq:volume_conservation})) or improved IBM schemes \citep{wu2012simulation,casquero2020divergence}. Note the new fluid velocity $\mathbf{u}(\mathbf{x},t+\Delta t)$ is obtained after solving the discretized lattice Boltzmann equation (\ref{eq:lbe}) which requires an {\it a priori} knowledge of the membrane forces. The membrane forces $\mathbf{F}_k^{m}(t)$ are computed before solving equation (\ref{eq:lbe}), so to say, with respect to the old position of the membrane nodes $\mathbf{X}^m(t)$. Thus, the following notation $\mathbf{u}(t+\Delta t)$ is used in equation (\ref{eq:ibm_velocity_interpolation}) instead of $\mathbf{u}(t)$.
In discrete forms, equations (\ref{eq:ibm_force_spreading}) and (\ref{eq:ibm_velocity_interpolation}) can be rewritten such that
\begin{align}
     & \mathbf{f}(\mathbf{x},t) = \sum_k \mathbf{F}_k^{m}(t) \delta(\mathbf{x}-\mathbf{X}_k^{m}(t)) \Delta \Omega_{S,k} \\
     & \mathbf{u}(\mathbf{X}^{m},t+\Delta t) = \sum_{\mathbf{x}}  \mathbf{u}(\mathbf{x},t+\Delta t) \delta(\mathbf{x}-\mathbf{X}^{m}(t)) \Delta x^3,
\end{align}
where $\sum_k$ runs over the membrane nodes located within the interpolation range of a given fluid node $\mathbf{x}$, and $\sum_{\mathbf{x}}$ over a cuboidal region centred around a given membrane node $\mathbf{X}^m$.
The advection equation of a Lagrangian node on the capsule follows an explicit forward Euler scheme, such that
\begin{equation}
    \mathbf{X}^{m}(\xi_1,\xi_2,t+\Delta t) = \mathbf{X}^{m}(\xi_1,\xi_2,t) +  \mathbf{u}(\mathbf{X}^{m},t+\Delta t)\Delta t.
    \label{eq:ibm_euler_forward}
\end{equation}
Let us remark that, in principle, depending  on the specific problem, one may need to tune the time step of the numerical integration of the Lagrangian dynamics independently from the lattice Boltzmann time step. For instance, for vanishing values of $C$, mesh instabilities can arise (owing to the formation of large wrinkles on the capsule surface). We have, therefore, decided to keep $C \ge 10^{-3}$ such that choosing the time-step equal to the lattice Boltzmann $\Delta t$ proved sufficient to prevent such instabilities.

The Dirac delta function in equations (\ref{eq:ibm_force_spreading}) and (\ref{eq:ibm_velocity_interpolation}) is usually replaced with a smoother interpolation function ($\varphi$) of some shape such as $\delta(\mathbf{x}) = \varphi(x_1)\varphi(x_2)\varphi(x_3)$ to avoid jumps in velocities or in the applied forces occurring when the Lagrangian nodes do not coincide with the nodes of the Eulerian grid.
Several distribution functions have been used in the IBM literature for a wide range of applications. For detailed reviews on the IBM and its accuracy, we refer the reader to \cite{mittal2005immersed} and \cite{kruger2017lattice} among other existing works on this topic.  
In what follows, we use a two-point linear interpolation function as discussed in \cite{kruger2012computer}, which is given by
\begin{equation}
    \varphi(\hat{x}) = \left\{
    \begin{array}{rcr}
    & 1 - |\hat{x}| \quad \text{for} \quad 0 \le |\hat{x}| \le 1 \\
    & 0 \quad \text{for} \quad |\hat{x}| \ge 1
    \end{array}
    \right. ,
\end{equation}
where $\hat{x}$ can denote $x_1$, $x_2$, or $x_3$. 

\subsection{Simulations details, key parameters and observables}
We simulate shear flows, with constant shear rate $\dot{\gamma}$, in cubic boxes, seeded with $N_p$
capsules of radius $r$.
The computational domain has a side length $L=128\Delta x=16r$ and it is biperiodic along the flow and vorticity directions (respectively $x_1$ and $x_2$ directions).
It is bounded, in the $x_3$ direction, by two planar walls at which we impose a velocity boundary condition such to generate the driving shear flow as described in \cite{hecht2010implementation}.\\
The main control parameters of the problem are the volume fraction of capsules, 
$\phi=\frac{4 \pi r^3}{3L^3}$, the capillary number, $Ca$, and the membrane inextensibility, 
$C$ (the particle scale Reynolds number being
always so small, $R_e = (\rho_0\dot{\gamma} r^2)/\mu_0 < 10^{-1}$, that the dynamics can be considered effectively inertia-less).
The capillary number, quantifying the relative intensity of viscous and elastic forces, is defined as
\begin{equation}
  Ca = \dot{\gamma} \tau_{el},
\end{equation}
where $\tau_{el} = (\mu_0 r) / G_s$ is a time scale associated to the elasticity of the capsule. The value of $C$ depends on the composition of the membrane. For example albumin–alginate capsules have a $C$ of the order of unity \citep{carin2003}, while for red blood cells $C \gg 1$.

Following \cite{batchelor1970stress}, we evaluate the average particle stress tensor as:
\begin{equation}
  \Sigma_{ij}^p = \frac{1}{N} \sum_{\alpha=1}^{N} n S^\alpha_{ij} = -\frac{1}{V_D} \sum_{\alpha=1}^{N} \int_{\Omega_S} \frac{1}{2}\{F_i^{m,\alpha} X_j^{m,\alpha} + F_{j}^{m,\alpha} X_i^{m,\alpha}\} d\Omega_S^{\alpha},
\end{equation}
where $i$ and $j$ are indices referring to Cartesian directions, $\sum_{\alpha=1}^{N_p}$ is a sum over the number of particles in the averaging volume $V_D$, $n$ the number density, and $\mathbf{S}$ the particle stresslet. 
Here $d\Omega_S^{\alpha}$ is the area element centred at $\mathbf{X}^{m}$, and $\mathbf{F}^{m}$ is the surface force density exerted by the membrane on the fluid.
The rheology of the system is then assessed in terms of the suspension relative viscosity 
and normal stress differences. 
The relative viscosity of the suspension $\mu_r$ is defined as:
\begin{equation}
   \mu_r = \frac{\mu}{\mu_0} = 1 + \frac{\Sigma_{13}^p}{\mu_0\dot{\gamma}}.
\end{equation}
where $\mu$ is the effective viscosity of the system. 
The first and second normal stress differences can be deduced from the average particle stress 
tensor as:
\begin{equation}
  N_1 =  \Sigma_{11}^p - \Sigma_{33}^p, \quad
  N_2 =  \Sigma_{33}^p - \Sigma_{22}^p.
\end{equation}
The deformation of a capsule in the shear plane can be characterised 
in terms of the Taylor deformation parameter \citep{taylor1934formation}
\begin{equation} \label{eq:taylor}
D = \frac{r_1 - r_3}{r_1 + r_3},
\end{equation}
where $r_1$ and $r_3$ are the major and minor principal semi-axes (in the shear plane) of an ellipsoid having the same 
tensor of inertia as the deformed capsule,
and the inclination angle $\theta$, which is the angle the major axis forms with the positive direction of the $x_1$-coordinate (see figure \ref{fig:01}).
The ellipsoid principal semi-axes are defined as \citep{ramanujan1998deformation,li2008front,frijters12,farutin20143d}: 
\begin{equation}
r_1 = \sqrt{\frac{5}{2\rho_0 V}(I_2 + I_3 - I_1)}, \quad
r_2 = \sqrt{\frac{5}{2\rho_0 V}(I_1 + I_3 - I_2)}, \quad
r_3 = \sqrt{\frac{5}{2\rho_0 V}(I_1 + I_2 - I_3)},
\end{equation} 
where $I_1$, $I_2$, and $I_3$ are the eigenvalues of the tensor of inertia.

The rheology and microstructure of suspensions of strain-hardening capsules up to a volume fraction of $0.5$ are studied for capillary numbers ranging from $Ca=0.1$ to $1$.  The number of particles is varied from $1$ to $500$, corresponding to $\phi\approx 0.001$ and $\phi \approx 0.5$, respectively. Each particle is discretized with $1280$ triangles and $642$ nodes, and initialized as a sphere with a radius $r=8\Delta x$. When the distance between two neighbouring particles is below $1\Delta x$, a repulsive force acts on the surface of the two particles with an interaction strength chosen as $\bar{\epsilon} \approx 10^2 G_s r$.
The simulations are initialised with the capsules distributed randomly in the domain with an initial radius $r_0 < r$.
The radius of each capsule is then increased in time with a fixed growth rate until reaching the desired size.
The relative error on the capsule's volume defined as $\epsilon_V = |V-V_0|/V_0$ is below $0.03\%$ in all simulations.
For suspensions, the measured quantities are obtained from an average over the number of particles and over time. 
In terms of strain units ($\dot{\gamma}t$), our simulations reach convergence after the first $5$ to $7\dot{\gamma}t$. The time average is performed after the initial transient state defined here as the first $10\dot{\gamma}t$. Time histories of the mean deformation and relative viscosity in the dilute and semi-dilute limits, together with the effect of mesh discretization and finite size effects, are shown in the Appendix. Details on the performance of our code can be found in \cite{aouane2018mesoscale}.\\

\section{\label{sec:results}Results}

\subsection{Behaviour of a single capsule in a shear flow}\label{subsec:singlepart}
\begin{figure}
	\centering
	\includegraphics[width = 1.\textwidth]{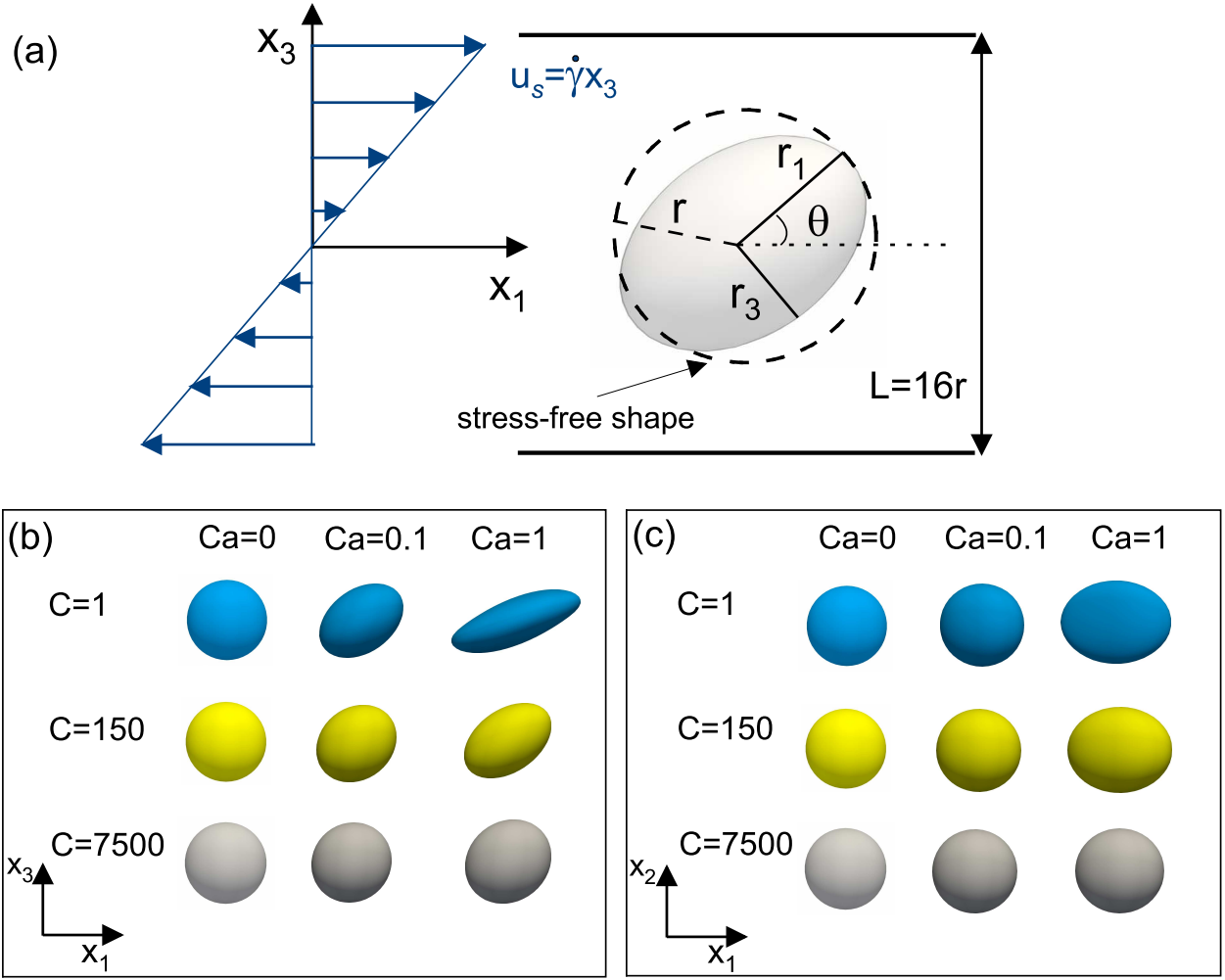}
	\caption{(a) Schematic of a single capsule in a shear flow showing the initial (dashed lines) and the typical ellipsoidal steady-state shapes; $r$ is the radius of the unstressed sphere, $r_1$ and $r_3$ are the major and minor semi-axes in the shear plane
	($r_2$ is the one in the vorticity direction, not shown) and $\theta$ is the inclination angle. (b) and (c) Numerically computed steady-state shapes for different values of $C$ and $Ca$ namely in the {\it $x_3x_1$}-plane (defined between the shear gradient and flow directions) and {\it $x_2x_1$}-plane (defined between the vorticity and the flow directions). } 
	\label{fig:01}
\end{figure}

\begin{figure}
        \centering
        \includegraphics[width = 1.\textwidth]{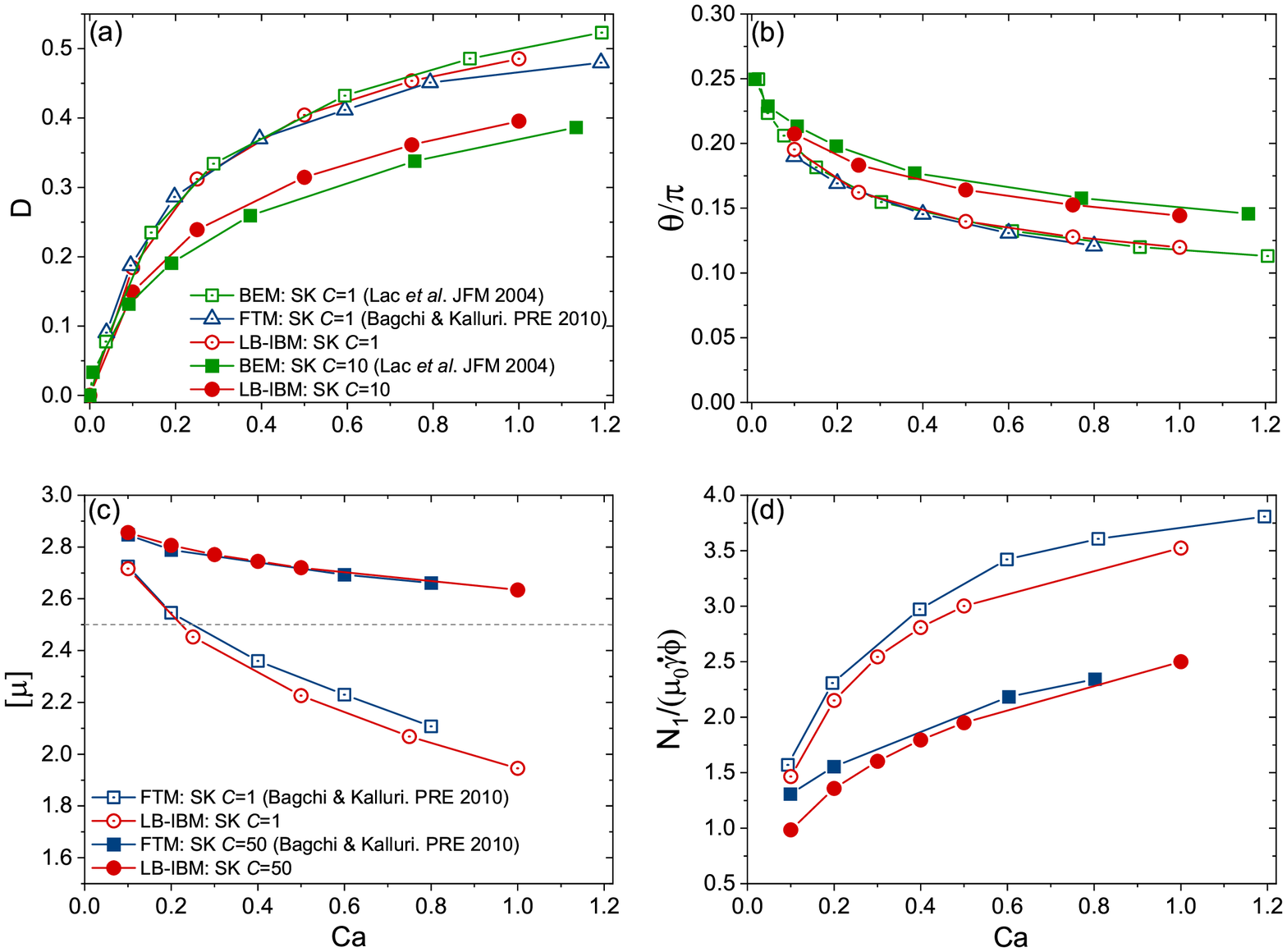}
        \caption{Steady Taylor deformation parameter (a), inclination angle (b), intrinsic viscosity (c), and first normal stress difference (d) of a single capsule under a shear flow. The open and full symbols in (a) and (b) are for $C=1$ and $C=10$, respectively, and for $C=1$ and $C=50$ in (c) and (d). BEM denotes the boundary element simulations of \cite{lac2004spherical}, and FTM the front-tracking method of \cite{bagchi2010rheology}. LBM-IBM are the numerical results obtained using our lattice Boltzmann code. The dashed line in (c) indicates the value of the intrinsic viscosity of a dilute suspension of rigid spheres. The legends for (b) and (d) are indicated in (a) and (c), respectively.
        }
        \label{fig:02} 
\end{figure}

\begin{figure}
        \centering
        \includegraphics[width = 1.\textwidth]{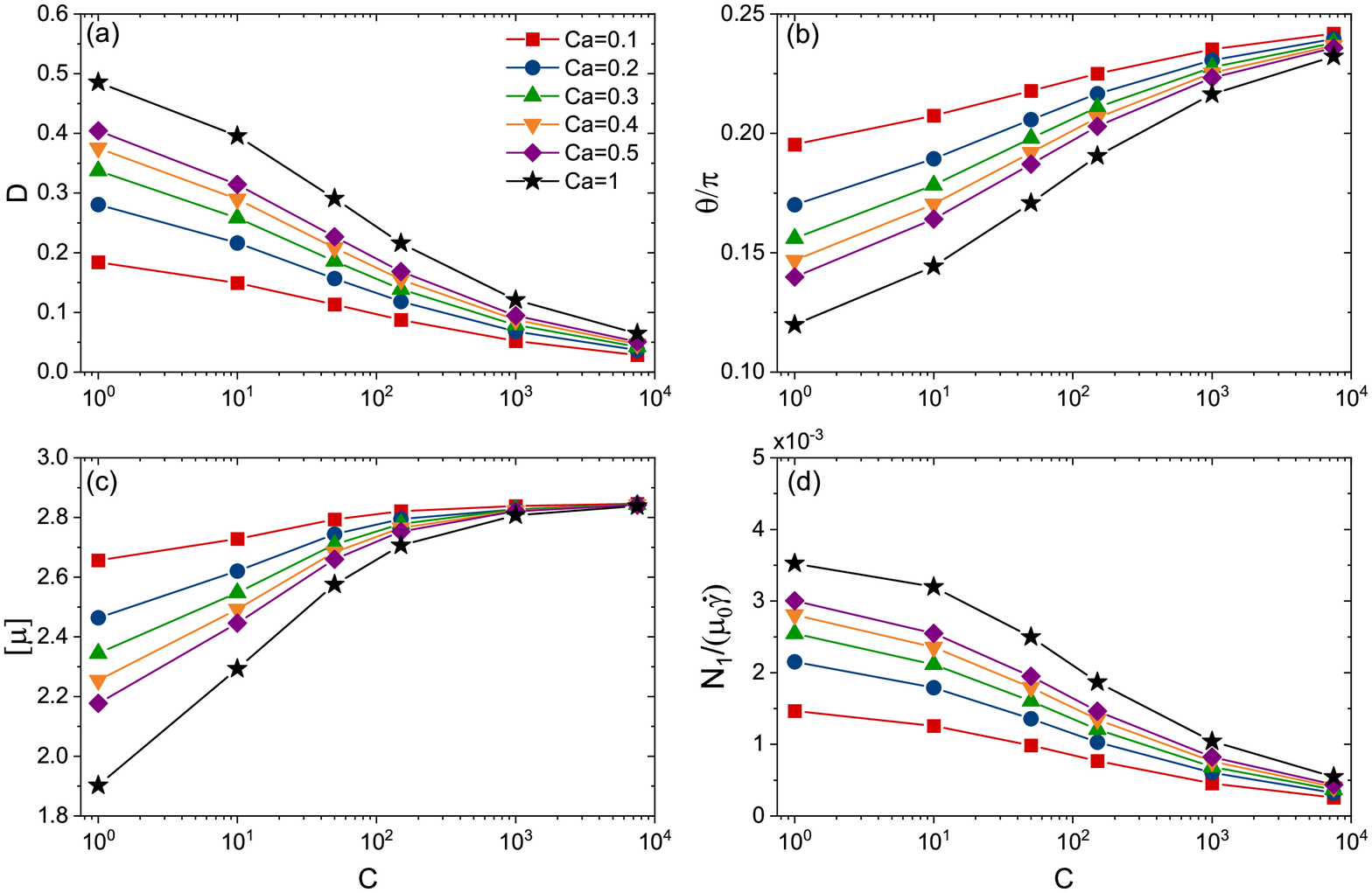}
        \caption{Effect of the membrane area incompressibility on the steady-state Taylor deformation parameter (a), inclination angle (b), intrinsic viscosity (c), and first normal stress difference (d) of a single Skalak capsule under a shear flow for different capillary numbers.
        The legend is given in (a).}
        \label{fig:03}
\end{figure}

In order to validate our approach, we first limit our simulations to the case of single initially spherical Skalak capsules. 
Our simulation domain is bounded by two parallel walls and the confinement is set to $\chi=2r/L=0.125$, so that the effect of the boundaries can be neglected.
A schematic representation of the simulation setup, together with examples of steady state shapes for different $Ca$ and $C$, are depicted in Fig.~\ref{fig:01}.
The steady Taylor deformation parameter, the inclination angle, the normal stress difference and the intrinsic viscosity, defined as
\begin{equation}
    [\mu] = \lim_{\phi \rightarrow 0} \frac{\mu_r - 1}{\phi},
\end{equation}
as functions of $Ca$ (for different values of $C$) are plotted in Fig.~\ref{fig:02}. We compare with numerical results obtained using the boundary element method (\cite{lac2004spherical}) and the front-tracking method (\cite{bagchi2010rheology}), showing good agreement.\\ 
We next move to explore systematically the parameter space spanned by $(Ca,C)$, with $Ca \in [0.1; 1]$ and $C \in [1; 7500]$. The corresponding data on steady-state elongation, inclination angle, intrinsic viscosity and first normal stress difference are reported in Fig.~\ref{fig:03}.
We see, from the symbols in Fig.~\ref{fig:03}(c), at fixed $C$, that $[\mu]$ decreases with $Ca$, denoting a shear-thinning character. The latter, in turn, appears to be directly correlated to an increase of the elongation of the capsule (Fig.~\ref{fig:03}(a)) and a decrease of its orientation with respect to the flow direction (Fig.~\ref{fig:03}(b)), similarly to what has been reported for drops, vesicles, and strain-softening capsules. 
Moreover, a clear effect of $C$ on the various observables can be detected:
as $C$ grows, their dependence on $Ca$ gets weaker. In particular, the steady Taylor parameter $D$ suggests that the particle is less and less deformed, i.e.~it approaches the limit of a rigid sphere. Correspondingly, $[\mu]$ varies less and less with $Ca$ and eventually the shear-thinning is suppressed.
For sufficiently large $C$, then, a very dilute suspension of strain-hardening capsules tends to behave rheologically as a suspension of rigid spheres (notice also that $N_1$ tends to zero, Fig.~\ref{fig:03}(d)), albeit with an intrinsic viscosity surprisingly slightly larger than the Einstein coefficient $[\mu]=5/2$ 
(for $C \gg 1$, $[\mu]$ seems to approach the limiting value $\approx 2.84$; see also \cite{bagchi2010rheology} for comparison).

\subsection{Suspensions: structure}
\begin{figure}
        \centering
        \includegraphics[width = .95\textwidth]{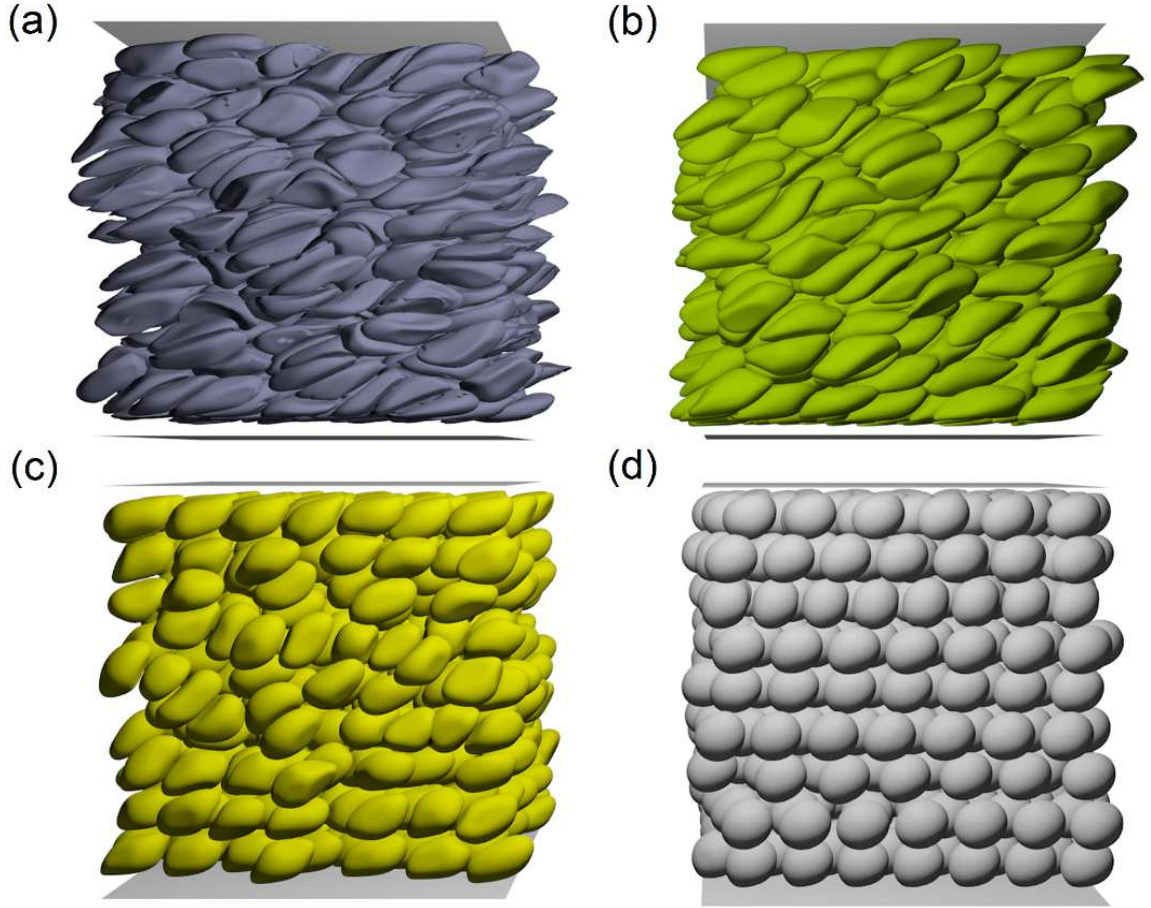}
        \caption{Steady-state configurations for $\phi=0.5$, $Ca=1$ and $C=0.1$ (a), $C=10$ (b), $C=1.5 \times 10^2$ (c) and 
        $C=7.5 \times 10^3$ (d).}
        \label{fig:04}
\end{figure}

\begin{figure}
        \centering
        \includegraphics[width = 1.\textwidth]{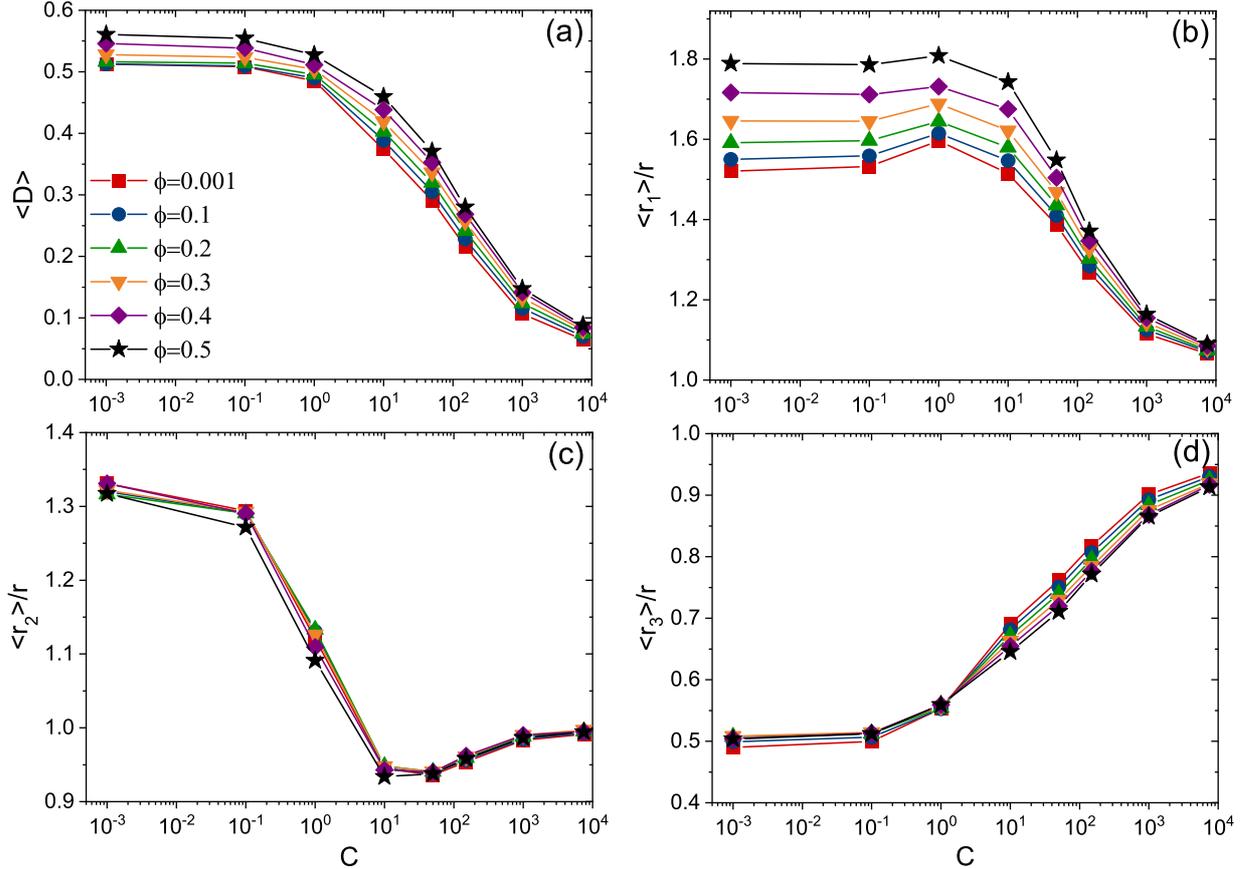}
        \caption{Steady-state Taylor deformation parameter (a). 
          Steady-state mean semi-axes of the capsules normalized by the reference radius as function of $C$ for
          different values of $\phi$ (b-d). The legend is indicated in (a). $Ca=1$.}
        \label{fig:05}
\end{figure}

We investigate the multi-particle case and the dependence of particle
shape and suspension rheological properties on the parameters describing the system,
namely $Ca$, $C$, and $\phi$. 
Examples of steady-state configurations of the suspension are shown in Fig.~\ref{fig:04},
for fixed $Ca=1$, $\phi=0.5$ and for four different values of $C=0.1,10,150,7500$.\\
In this subsection we focus on particle morphologies, characterized
in terms of the Taylor deformation parameter and the semi-axes of the equivalent
ellipsoid, whereas in the next one we will study the rheological properties of
the suspensions, highlighting their relations with the structure.
Analogously to the single particle case, the mean Taylor parameter $\langle D \rangle$ of
the suspension decreases with increasing $C$ (Fig.~\ref{fig:05}(a)), with a steepest
descent for $1 < C < 10^3$, which confirms that capsules become less deformable and tend to resemble rigid particles.
We recall, though, that such a parameter contains information
only on two of the three semi-axes of the equivalent ellipsoid. To get a deeper insight,
then, we present all three of them separately in Fig.~\ref{fig:05}(b-d). 
Interestingly, a non-monotonic behaviour is found for $\langle r_1 \rangle$ and
$\langle r_2 \rangle$ (the semi-axes in the flow and vorticity directions, respectively)
for $0.1<C<10$. In particular, for decreasing $C<10$, $\langle r_2 \rangle$,
whose direction is orthogonal to the elongational one, grows.
Moreover, $\langle r_2 \rangle$ stays always larger than $\langle r_3 \rangle$,
indicating that particles, on average, are not spheroids (eventually, for very large
$C$ particles approach the undeformable limit and the quasi-spherical shape,
$\langle r_i \rangle \rightarrow r$ for $i=1,2,3$, is recovered). 
In this sense, capsules display a lower level of symmetry than droplets,
which is to be attributed to the non-linear elastic characteristics of the membrane. 
The behaviour, in fact, persists across the various volume fractions explored,
even for the lowest $\phi$ (corresponding to the single particle case),
suggesting that the effect originates from the properties of the single particle
stress tensor. The spread of the mean principal semi-axes, 
mostly of $\langle r_1 \rangle$, 
is such that the product $\prod_{i=1}^3 \langle r_i \rangle$ varies with the volume fraction.
Such dependence is related to the variances of the distributions of the principal semi-axes 
(given that the mean volume of capsules, $\langle \prod_{i=1}^3 r_i \rangle$ is conserved)
and reflects, therefore, the spread in sizes, which decreases as $C$ increases,
  because the capsules become more and more rigid (in other words 
  the distribution tends to become sharply peaked around 
  $r_1 \sim r_2 \sim r_3 \sim r$).

Next, we consider how the particle deformation depends on the applied load,
for given material properties. It is tempting, first of all, to investigate how the peculiar behaviour for small/moderate $C$'s shows up across different shear values. 
	\begin{figure}
		\centering 
		\includegraphics[width = 1.\textwidth]{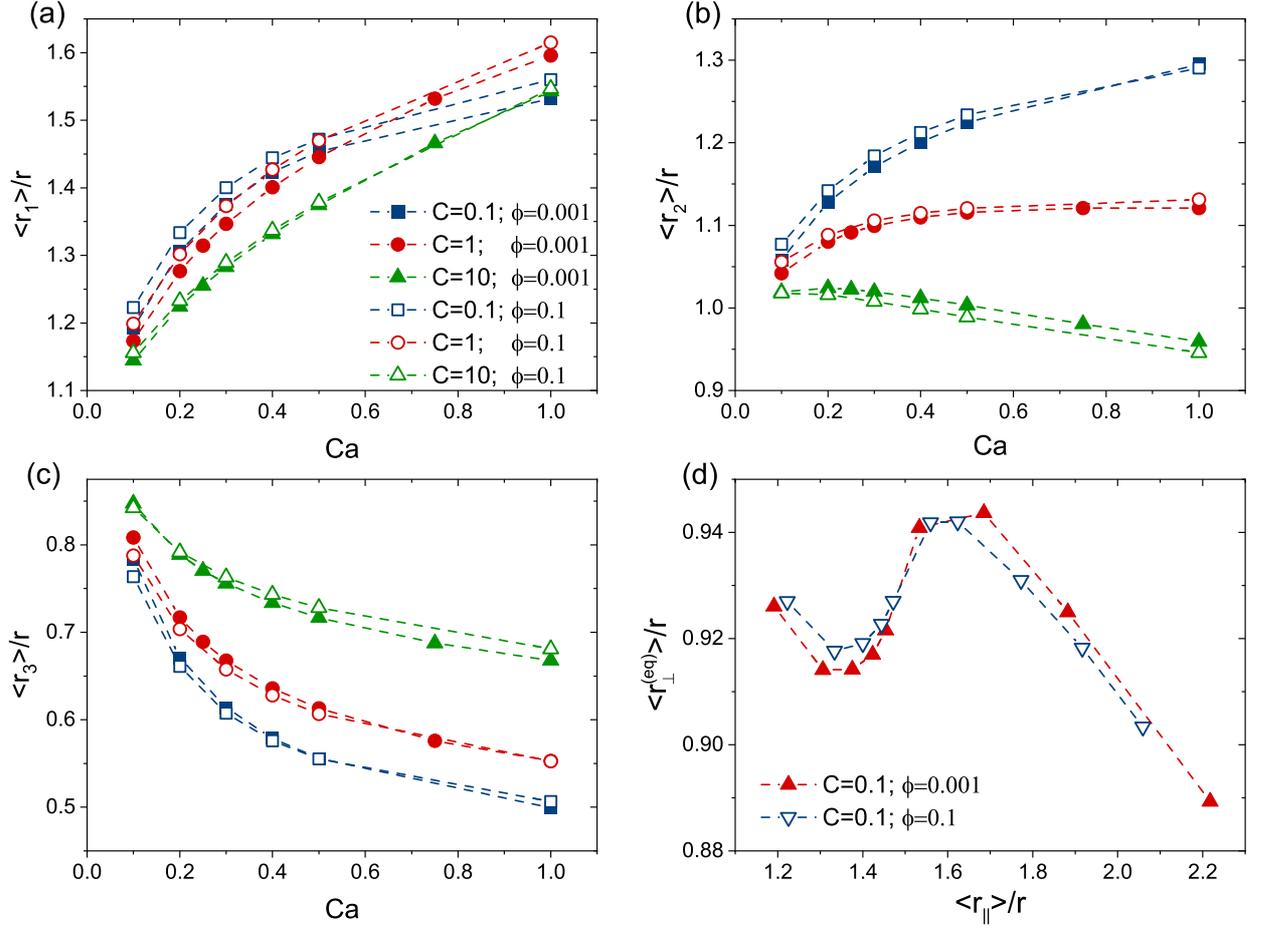}
		\caption{Steady-state semi-axes (normalized by the reference radius of the initially spherical stress-free capsule) as a function of $Ca$ for capsules with $C=0.1,1,10$ (a-c) (the legend is indicated in (a)). Transversal deformation versus elongation (normalized by the reference radius) for capsules with $C=0.1$, and $Ca \in [0.1; 7.5]$ (d). Open symbols refer to $\phi=0.1$ (semi-dilute suspension), while full symbols to $\phi=0.001$ (dilute suspension).}
		\label{fig:06}
	\end{figure}
	In figure \ref{fig:06}(a-c) we report the variation of $r_i$ for single 
	capsules with $C=0.1,1,10$, as a function of $Ca$. As $Ca$ increases, the capsules get, obviously, more elongated in 
	the extensional flow direction ($r_1$ grows), but the two minor semi-axes display opposite trends, depending 
	on the value of $C$: while $r_3$,
	as expected, decreases (for all $C$), $r_2$ (the one aligned with the vorticity direction) grows for $C<1$. 
	For the smallest $C$ considered, $C=0.1$, we need, therefore, to find an 
	equivalent breadth, $r_{\perp}^{\mbox{\tiny{(eq)}}}$ quantifying the degree of shrinkage or expansion 
	of the capsule in the equatorial plane. 
	We define this $r_{\perp}^{\mbox{\tiny{(eq)}}}$ as the ratio of the length of the membrane cross-section 
	(an ellipse) over $2\pi$ (such that it would be precisely the minor axis, if the capsule were a prolate spheroid), i.e.
	$r_{\perp}^{\mbox{\tiny{(eq)}}} = \frac{4 r_2 \mathcal{E}(\epsilon(r_2,r_3))}{2 \pi}$, 
	where $\mathcal{E}(x)$ is the complete 
	elliptic integral of the second kind (\cite{AbramowitzStegun}) and 
	$\epsilon(r_2,r_3) = \sqrt{1-\left(\frac{r_3}{r_2}\right)^2}$ is the eccentricity of the ellipse.
	In figure \ref{fig:06}(d), we plot the transversal deformation, represented by $r_{\perp}^{\mbox{\tiny{(eq)}}}$ versus
	the elongation, $r_{||} \equiv r_1$, both normalised by the rest radius $r$, for the capsule with $C=0.1$. It can be seen that, although 
	$r_{\perp}^{\mbox{\tiny{(eq)}}}/r$ never exceeds $1$, i.e. overall the membrane cross-section shrinks
	with respect to the equilibrium shape, there is a range of $Ca$ for which it expands as the capsule is elongated.
	This is an intriguing behaviour, in fact the opposite of the slope of the curve in Fig.~\ref{fig:06}(d),
	$\tilde{\nu}_s = -\frac{d r_{\perp}^{\mbox{\tiny{(eq)}}}}{d r_{||}}$, can be interpreted
	as a {\it local} Poisson's ratio, which is negative for 
	$1.3 r \stackrel{<}{\sim} r_{||} \stackrel{<}{\sim} 1.7 r$. This observation hints at a sort of local 
	(in shear) ``auxeticity'' (\cite{EvansEtAl}) of membranes obeying a Skalak-type constitutive law with low values of the 
	membrane inextensibility parameter.

\begin{figure}
        \centering
        \includegraphics[width = 0.75\textwidth]{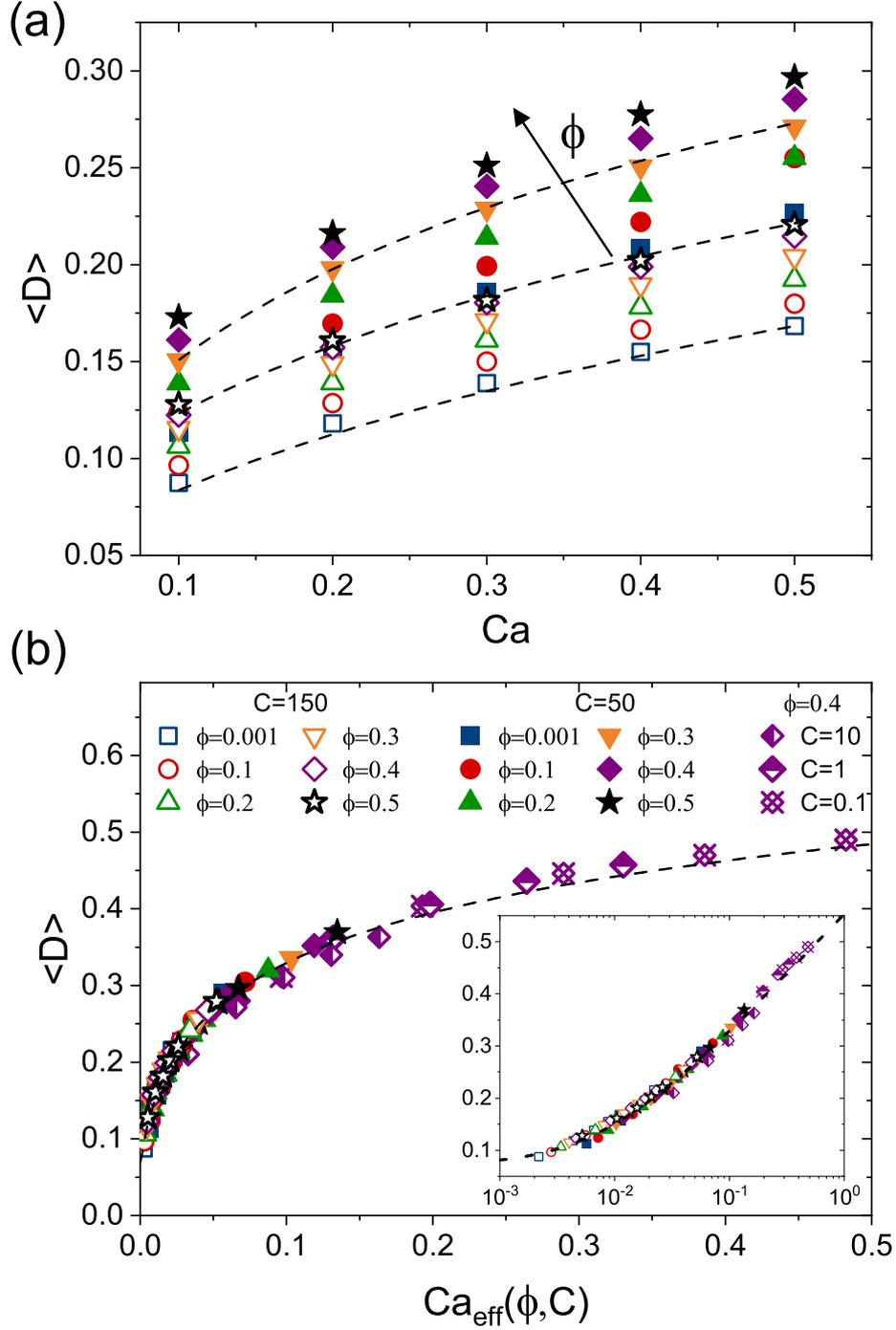}
        \caption{
          (a) Mean Taylor deformation parameter for a suspension of strain-hardening capsules as a function of $Ca$, for two values of membrane inextensibility, $C=50$ (filled symbols) and $C=150$ (open symbols). The dashed lines are fits of the numerical data using equation (\ref{eq:Dfit}) with $A=0.1$ and (from bottom to top) $Ca^*=0.13$, $Ca^*=0.07$ and $Ca^*=0.037$, respectively. The arrow indicates a growing volume fraction $\phi$. (b) Mean Taylor deformation parameter as a function of the effective capillary number, Eq. (\ref{eq:Caeff}). The dashed line corresponds to the fitting function (\ref{eq:Dfit}) ($A_{\mbox{\tiny{eff}}}=0.1$ and $Ca_{\mbox{\tiny{eff}}}^*=4 \times 10^{-3}$). Inset: Lin-Log plot of $\langle D \rangle$ vs $Ca_{\mbox{\tiny{eff}}}$, highlighting the logarithmic behaviour for $Ca_{\mbox{\tiny{eff}}}>Ca_{\mbox{\tiny{eff}}}^*$. The legend is indicated in (b).
               }
        \label{fig:07}
\end{figure}

In figure \ref{fig:07}(a) we show the average Taylor deformation parameter as function of $Ca$, for various $\phi$'s. Two sets of data corresponding to $C=50$ (closed symbols) and $C=150$ (open symbols) are reported.
The deformation grows as $\langle D \rangle \sim Ca$ for small capillary numbers, as expected, and then sub-linearly as the $Ca$ increases (eventually we observe a logarithmic dependence $\langle D \rangle \sim \log(Ca)$, in agreement with previous numerical \citep{DodsonDimitrakopoulos} and experimental \citep{HardemanEtAl} findings), reflecting the strain-hardening character of the capsules.
It can be asked whether one may find a functional form that allows to recast the variability among curves into a single curve shape parameter, $Ca^*(\phi,C)$, that is
\begin{equation}\label{eq:Dfit}
\langle D \rangle \equiv \mathcal{D}(Ca,\phi,C) = Ag\left(\frac{Ca}{Ca^*(\phi,C)}\right),
\end{equation}
where $A$ is a constant prefactor and the function $g(x)$ has to  be chosen such that it reproduces and connects both behaviours at small and large $Ca$, that is $g(x) \sim x$ as $x \rightarrow 0$ and $g(x) \sim \log(x)$ for $x \gg 1$. This is indeed possible and we show it in figure \ref{fig:07}(a). There, the fits, indicated by the dashed lines, are obtained from equation (\ref{eq:Dfit}), choosing for the function $g$ the expression $g(x) = \log(1+x)$, with the same $A=0.1$ and, from bottom to top, $Ca^* = 0.13$, $Ca^*=0.07$, and $Ca^*=0.037$, respectively. For a fixed capillary number, $\langle D \rangle$ increases linearly with the volume fraction $\phi$, similarly to suspensions of drops and neo-Hookean capsules \citep{loewenberg1996numerical,matsunaga2016rheology}. Conversely, the larger is the membrane inextensibility $C$, the less deformed are the capsules.

Given the self-similar form of (\ref{eq:Dfit}), we would like to find a 
universal curve for $\langle D \rangle$, through a proper definition of an effective capillary number $Ca_{\mbox{\tiny{eff}}}$.
The enhancement of deformation with $\phi$ can be interpreted as an effect
of larger viscous stresses around the particle, due to the fact that the effective
viscosity of the suspension increases with the volume fraction (this aspect will
be discussed in more detail in the next subsection).
This suggests that we should replace, in $Ca_{\mbox{\tiny{eff}}}$, the ``bare'' dynamic viscosity with the
effective one, $\mu_0 \rightarrow \mu_{\mbox{\tiny{eff}}} = \mu_0(1+[\mu]\phi)$. 
Here we assume, for simplicity, linearity in $\phi$ and a constant (with $Ca$ and $C$)
intrinsic viscosity, equal to its large $C$ limit, $[\mu] \approx 2.8$
(see section \ref{subsec:singlepart}).

Furthermore, we note that for a non-zero membrane inextensibility, it is more appropriate to base the capillary number on the Young's modulus instead of the shear modulus (\cite{barthes1981time}). We propose to replace $G_s$ with $E_s=\frac{(2+4C)}{(1+C)}G_s$. However, this might not be sufficient. In fact, the imposed constraint of volume conservation, for a spherical equilibrium shape, effectively entails an extra-tension on the surface, since the capsules tend to become essentially undeformable as the area dilatation modulus is increased. This can be accounted for by the substitution
	\begin{equation}
	E_s \rightarrow E_s^{\mbox{\tiny{(eff)}}}=\frac{(2+4C)}{(1+C)}G_s^{\mbox{\tiny{(eff)}}}=\frac{(2+4C)}{(1+C)}(1+\beta C)G_s
	\end{equation}
	($\beta$ being a free
	parameter, which we set hereafter to $\alpha=0.07$) in
	the effective capillary number, which then finally reads
	\begin{equation}\label{eq:Caeff}
	Ca_{\mbox{\tiny{eff}}}(\phi,C) =
	\frac{\mu_{\mbox{\tiny{eff}}}\dot{\gamma}r}{E_s^{\mbox{\tiny{(eff)}}}}=
	\frac{(1+[\mu]\phi)(1+C)}{(2+4C)(1+\beta C)}\left(\frac{\mu_0\dot{\gamma}r}{G_s}\right)
	\equiv \frac{(1+[\mu]\phi)(1+C)}{(2+4C)(1+\beta C)}Ca.
	\end{equation}
When plotted as function of $Ca_{\text{\tiny{eff}}}$, the values of 
$\langle D \rangle$ for different $\phi$ and $C$ collapse onto a single master curve,
as shown in figure \ref{fig:07}(b). Such curve can be also fitted using (\ref{eq:Dfit}),
with $A=0.1$ and $Ca_{\mbox{\tiny{eff}}}^*=4 \times 10^{-3}$.
Notice that the existence of a single $Ca_{\mbox{\tiny{eff}}}^*$ capable to fit all data sets upon the rescaling (\ref{eq:Caeff}) is equivalent to say that the dependence of the curve-shape parameter $Ca^*$ on $Ca$ and $C$ 
is such that $Ca^* \propto \frac{G_s^{\mbox{\tiny{(eff)}}}(C)}{\mu_{\mbox{\tiny{eff}}}(\phi)}$.

\subsection{Suspensions: rheology}

\begin{figure}
        \centering
        \includegraphics[width = 1\textwidth]{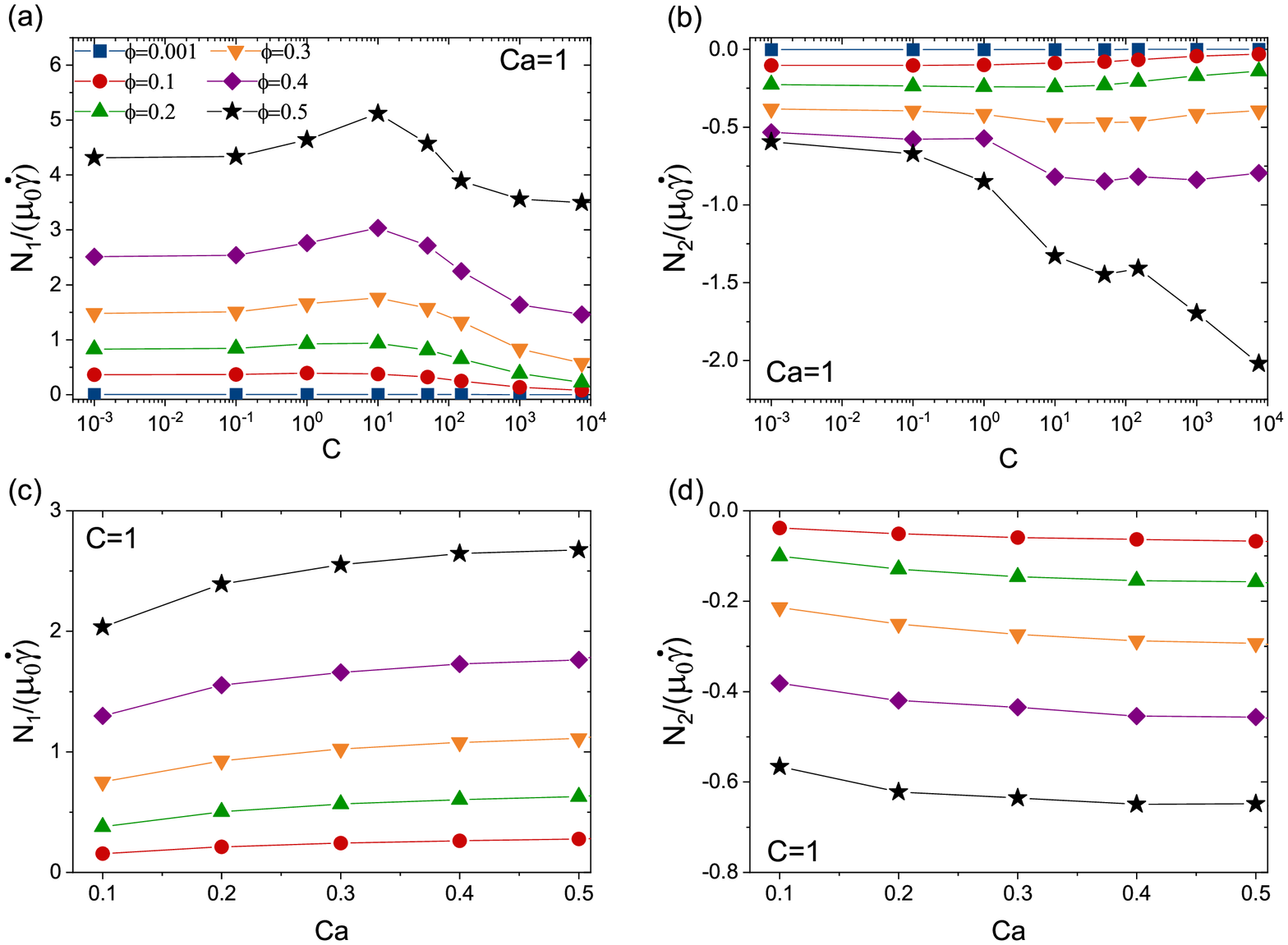}
        \caption{Normal stress differences (normalised by the dynamics viscosity of the fluid 
        times the applied shear) as function of $C$ for $Ca=1$ (a,b), and as function of $Ca$ for $C=1$ (c,d). The legend is shown in (a).}
        \label{fig:08}
\end{figure}

We now consider the rheological response of the system
by looking at the suspension relative viscosity and normal stress differences.
In figure \ref{fig:08}, we plot $N_1$ and $N_2$ 
(normalised by $\mu_0 \dot{\gamma}$), 
for various volume fractions, for $Ca=1$ as a function of the membrane inextensibility (panels (a) and (b)) and for $C=1$ as a function of the capillary number (panels (c) and (d)).
We see that both (though $N_2$ only weakly)
show a non-monotonic dependence on $C$, a behaviour that is enhanced as $\phi$ is increased. 
In particular, $N_1$ initially grows with $C$, reaches a peak at $C \approx 10$, and
then starts to decrease.
For emulsions and strain-softening capsules, the magnitude of $N_1$ is significantly larger than the magnitude of $N_2$, whereas the opposite is true for suspensions of rigid particles. $N_1$ and $N_2$ are known to be correlated to hydrodynamic interaction, and particle-particle collisions, respectively (\cite{guazzelli2018rheology}). 
We find that for strain-hardening capsules $N_1$ is positive and grows monotonically with $Ca$, whereas $N_2$ has a negative sign and decreases with $Ca$
(in qualitative agreement with what found for suspensions of strain-softening capsules 
\citep{matsunaga2016rheology}).
The magnitude of $N_1$ is always larger than $N_2$, but the ratio $|N_1|/|N_2|$ 
diminishes with the increase of $C$.  
It seems, therefore, in principle possible, by tuning their deformability through $C$, to make collections of such soft particles behave rheologically more as solid suspensions or as 
emulsions and suspensions of strain-softening capsules. \\

\begin{figure}
        \centering
        \includegraphics[width = 0.75\textwidth]{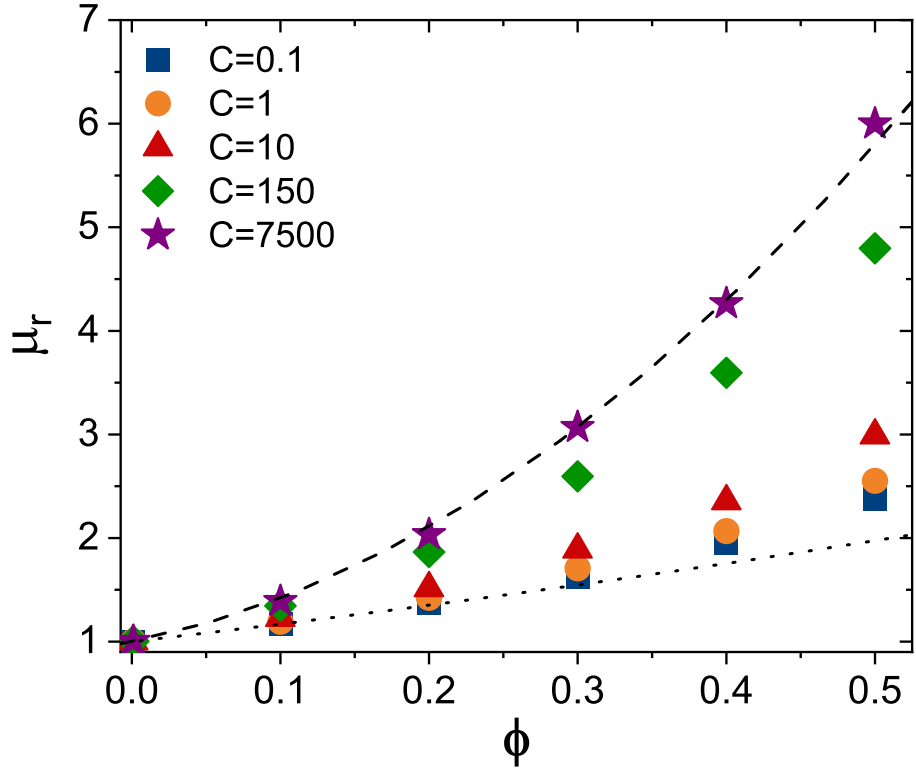}
        \caption{Relative viscosity as function of the volume fraction for several values of $C$ and $Ca=0.5$. The curves correspond to equation (\ref{eq:rel_visc_poly}) 
        with $[\mu]=1.63$, $K=0.64$ (dotted line) and $[\mu]=2.84$, $K=13.5$ (dashed line), respectively.}
        \label{fig:09}
\end{figure}

To delve deeper into this aspect, we study how the dependence of the relative viscosity of the suspension on the capsule volume fraction changes with $C$.
We report in figure \ref{fig:09} the behaviour of $\mu_r$ vs $\phi$ for $Ca=0.5$ 
and various $C$'s. The relative viscosity of a suspension can be expressed as a polynomial function of the volume fraction as
\begin{equation}
  \mu_r = 1 + [\mu]\phi + K \phi^2 + \mathcal{O}(\phi^3),  
  \label{eq:rel_visc_poly}
\end{equation}
where $[\mu]$ is the intrinsic viscosity and the second order term accounts for pair hydrodynamic interactions and allows to expand the validity of equation (\ref{eq:rel_visc_poly}) to semi-dilute cases \citep{batchelor1972determination}.
All data sets in figure \ref{fig:09} can be reasonably well fitted by a quadratic relation 
of the type (\ref{eq:rel_visc_poly}). As $C$ increases, the data tend to approach a limiting 
curve with $[\mu]=2.84$ and $K=13.5$ (dashed line), whereas for low $C$
they tend to agree well with the curve (dotted line) for strain-softening particles reported in \cite{matsunaga2016rheology}.\\
\begin{figure}
        \centering
        \includegraphics[width = 0.75\textwidth]{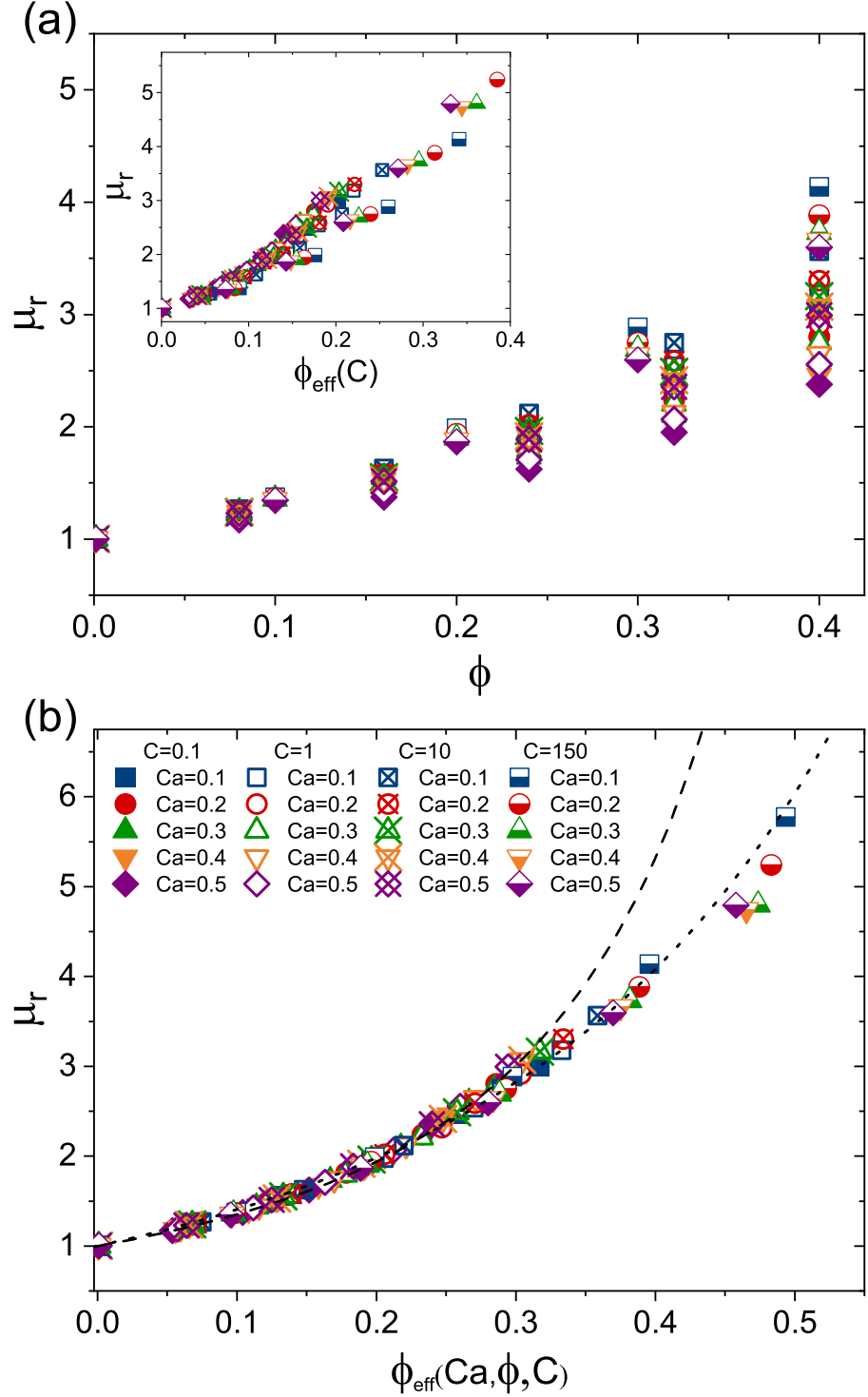}
        \caption{      	 	
          (a) Relative viscosity as a function of the volume fraction for different $Ca$ and $C$. The inset shows $\mu_r$ plotted as function of the effective volume fraction as defined in equation \eqref{eq:phieff}. The legend of (a) is shown in (b).
          (b) Relative viscosity as a function of the effective volume fraction (see equation \eqref{eq:phieff2}).
          The dashed line corresponds to a fit of the numerical data using equation \eqref{eq:fitEilers}, with $B=1.4$, and $\Phi_{\mbox{\tiny{m}}}=0.7$. Same for the dotted line but with $B=1.7$, and $\Phi_{\mbox{\tiny{m}}}=1.2$  
                }
        \label{fig:10}
\end{figure}
In figure \ref{fig:10}(a) we extend the study of the relative viscosity to a 
range of capillary numbers, $Ca \in [0.1, 0.5]$, in order to analyse the response
of the system to changes in the applied shear. 
The shear-thinning character of the suspension of capsules can be appreciated:
for a given volume fraction, in fact, $\mu_r$ tends to decrease with $Ca$
(the larger $\phi$ the more evident is the shear-thinning), 
but the spread is reduced for larger $C$, i.e. the shear-thinning tends to be suppressed,
confirming that also for semi-dilute and moderately concentrated regimes, the rheology of 
suspensions of strain-hardening capsules resemble that of solid suspensions.
In order to find a universal behaviour of the relative viscosity
across the various shear rates, 
recently \cite{rosti2018rheology} introduced, for suspensions of deformable viscoelastic spheres 
(and \cite{takeishiJFM2019} extended to the case of red blood cells),
the notion of a reduced {\it effective} volume fraction, calculated using for the particle volume
(when they are in the deformed state), that of a sphere with a radius equal to
the smallest particle semi-axis $r_3$, i.e.~$\Phi_{\mbox{\tiny{eff}}} = \frac{4}{3}\pi r_3^3$.
The rationale behind this approach is that the dynamically ``active'' 
direction is the velocity gradient (wall-normal direction) and, 
since the deformed particles do not tumble and tend to
align with the flow direction, then the relevant length is the minor axis.
For $r_3$, the average value as measured in the simulations was taken.
Here, we propose to relate $r_3$ to the radius at rest $r$ through the Taylor deformation
parameter $D$. To this aim, let us assume the particles to be prolate ellipsoids with
$r_1 > r_2 = r_3$ (this is an approximation that allows to close the problem, although
we know that the specific relation among the three axes depends on the value of membrane
inextensibility $C$).
$r_1$ and $r_3$ enter in the
expression of $D$ (equation (\ref{eq:taylor})), that can be inverted as
\begin{equation}
r_1 = \frac{1+D}{1-D}r_3.
\end{equation}
Due to volume conservation we have $r^3 = r_1r_3^2$ and therefore $r^3 = \frac{1+D}{1-D}r_3^3$,
which implies
\begin{equation}\label{eq:r3vsr}
r_3 = \left(\frac{1-D}{1+D}\right)^{1/3}r.
\end{equation}
If we now define, as in \cite{rosti2018rheology}, the effective volume fraction
as $\Phi_{\mbox{\tiny{eff}}} \equiv \frac{4}{3}\pi \langle r_3 \rangle^3 n$ and assume
(\ref{eq:r3vsr}) to be valid also for average quantities in the suspension, 
we get 
\begin{equation}\label{eq:phieff}
\Phi_{\mbox{\tiny{eff}}} = \frac{1-\langle D \rangle}{1+\langle D \rangle}\phi. 
\end{equation}
As we learned from the previous section, $\langle D \rangle$, in turn, depends on $\phi$, on
the capillary number and on $C$. If we plug equation (\ref{eq:Dfit}) inside (\ref{eq:phieff}),
with the rescaled effective capillary number, (\ref{eq:Caeff}), and we plot $\mu_r$ as a function of the obtained $\Phi_{\mbox{\tiny{eff}}}$, we observe a reasonable overlap of the data, although with some deviations, especially for the largest values of $C$ (see inset of Fig.~\ref{fig:10}(a)). 
We ascribe this partial failure to the fact that our strain-hardening capsules are not precisely aligned with the flow direction (see Fig.~\ref{fig:02}(b)). Consequently, the 
relevant, flow-orthogonal, length is not exactly $r_3$, but the vertical semi-axis of the 
ellipse, resulting as the section of the particle ellipsoid on a plane perpendicular to 
the flow direction (and crossing its centre). It is easy to show, by geometrical arguments, 
that such length is $\ell = r_3 \sqrt{\frac{1+b^2}{1+\left(\frac{r_3}{r_1}\right)^2b^2}}$,
where $b=\tan(\theta)$, and $\theta$ is the inclination angle. If we define the effective 
volume fraction in terms of this length (and recalling that $r_3/r_1 = (1-D)/(1+D)$), 
equation (\ref{eq:phieff}) becomes
\begin{equation}\label{eq:phieff2}
\Phi_{\mbox{\tiny{eff}}} = \frac{1-\langle D \rangle}{1+\langle D \rangle}
\left(\frac{1+b^2}{1+\left(\frac{1-\langle D \rangle}{1+ \langle D \rangle}\right)^2b^2}\right)^{3/2}\phi.
\end{equation}
The parameter $b$ itself depends, through $\theta$, on $Ca$ and $C$. However, for simplicity, we consider it here as a fit constant, taking for $\theta$ values restricted to the range 
in Fig.~\ref{fig:02}(b). In particular, for $\theta = \frac{\pi}{5}$ ($b\approx 0.726$), we get a nice collapse of all data sets onto a single master curve which can be well fitted, among others, with an Eilers function
(\cite{eilers1941viskositat})
\begin{equation}\label{eq:fitEilers} 
  \mu_r(\Phi_{\mbox{\tiny{eff}}}) = \left[1 + \frac{B\Phi_{\mbox{\tiny{eff}}}}{1-
    \frac{\Phi_{\mbox{\tiny{eff}}}}{\Phi_{\mbox{\tiny{m}}}}}\right]^2,
\end{equation}
with parameters $B=1.4$ and $\Phi_{\mbox{\tiny{m}}}=0.7$
(figure \ref{fig:10}(b)).
Choosing $B=1.7$ and $\Phi_{\mbox{\tiny{m}}}=1.2$ also the branch at large 
$\Phi_{\mbox{\tiny{eff}}}$ can be fitted, but these are somehow not sound values. 
We argue, instead, that the deviations from the Eilers fit (which occur for the set of data
corresponding to the largest $C=150$ and $\phi=0.4,0.5$) are due to the fact that under these conditions important hydrodynamic correlations emerge which cannot be simply adhered to a reduced volume effect. 
Let us stress that the relation (\ref{eq:phieff}), as much as in the approach of \cite{rosti2018rheology},
needs an empirical input (the function $g(x)$ in (\ref{eq:Dfit})). However, for small effective
capillary number, it is possible to approximate $g$ with its linear part, thus providing
a closed and explicit expression for $\Phi_{\mbox{\tiny{eff}}}(Ca,C,\phi)$ and, consequently, an explicit dependence of the relative viscosity $\mu_r$ on $Ca$, $C$ and $\phi$.

\section{\label{sec:conclusions}Conclusions}
The rheology of a suspension of strain-hardening capsules is investigated numerically from the 
dilute to the concentrated regimes in a simple shear flow. 
We have addressed the role of the membrane inextensibility, $C$,
on the capsule shape and on the suspension rheology, at varying the volume fraction, $\phi$, 
and the capillary number, $Ca$ (based on the applied shear).
Our results indicate that increasing the membrane inextensibility makes the capsules 
less deformable, and as a consequence the shear-thinning character of the suspension is hindered.
We show that, upon a proper definition of an effective $C$- and $\phi$-dependent capillary number, the mean Taylor deformation parameters relative to various data sets collapse
onto a single master curve. However, it proved necessary to go beyond the deformation parameter, in order to get a deeper insight on the complex impact the membrane inextensibility has on the full three-dimensional capsule shape. The three principal semi-axes displayed, in fact, a non-monotonic 
dependence on $C$, and for small $C$ ($C=0.1$) and a certain range of $Ca$, an auxetic behaviour of the capsules was observed.
The characteristic shape behaviour was reflected in a non-monotonic dependence of the first normal stress difference with the membrane inextensibility.
Finally, the rheological response of the suspension has been analysed also in terms of its relative viscosity. The latter showed a universal behaviour across the various concentrations, shear rates and membrane inextensibilities explored, once an effective volume fraction, taking into account the capsules elongation and orientation, was introduced.
\begin{acknowledgments}
{We are grateful to Marc L\'{e}onetti for fruitful discussions and to the referees for their comments and suggestions. 
The authors acknowledge financial support by the Deutsche Forschungsgemeinschaft (DFG) within the Cluster of Excellence ``Engineering of Advanced Materials'' (project EXC 315) (Bridge Funding), and within the research unit FOR2688 ``Instabilities, Bifurcations and Migration in Pulsatile Flows'' by the grant (HA4382/8-1). This work was also supported by the Competence Network for Scientific High-Performance Computing in Bavaria (KONWIHR III, project ``Dynamics of Complex Fluids''), and the European Cooperation in Science and Technology (COST) Action MP1305: ``Flowing Matter''.
The authors gratefully acknowledge the computing time granted by the John von Neumann Institute for Computing (NIC) and provided on the supercomputer JURECA at Jülich Supercomputing Centre (JSC), by the High Performance Computing Center Stuttgart (HLRS) on the Hazel Hen supercomputer, and by the Regional Computing Centre Erlangen (RRZE).}
\end{acknowledgments}

\appendix
\section{}{\label{sec:appendix}}

\subsection{Particle discretization and mesh quality}
Our membrane is discretized using triangular elements. The spherical capsule results from refining an icosahedron recursively $N_r$ times. The total number of faces denoted $N_f$ is defined from the total number of vertices $N_v$ and the number of recursive refinement such as $N_f=2N_v-4$, and $N_v=2+10 \cdot 4^{N_r}$. In this work, we used $N_r=3$ leading to $N_v=642$ and $N_f=1280$. To study the effect of the mesh discretization on the the shape and rheology, we varied $N_r$ from $2$ to $4$, which corresponds namely to $N_f=320$ and $N_f=5120$. We set $C$ to unity to test situations with large deformations. 

We report in table \ref{tab:mesh_quality} some of the relevant quantities measured for different number of faces $(N_f=320, 1280,5120)$, $C=1$ and $Ca=0.5$. The standard deviations of the different shape and rheology parameters are negligible for the three meshes, while the errors stemming from decreasing the number of faces of the mesh are significantly small.

\subsection{\label{sec:appendix-averages}Statistically stationary state and time averages.}

\begin{figure}
        \centering
        \includegraphics[width = 0.7\textwidth]{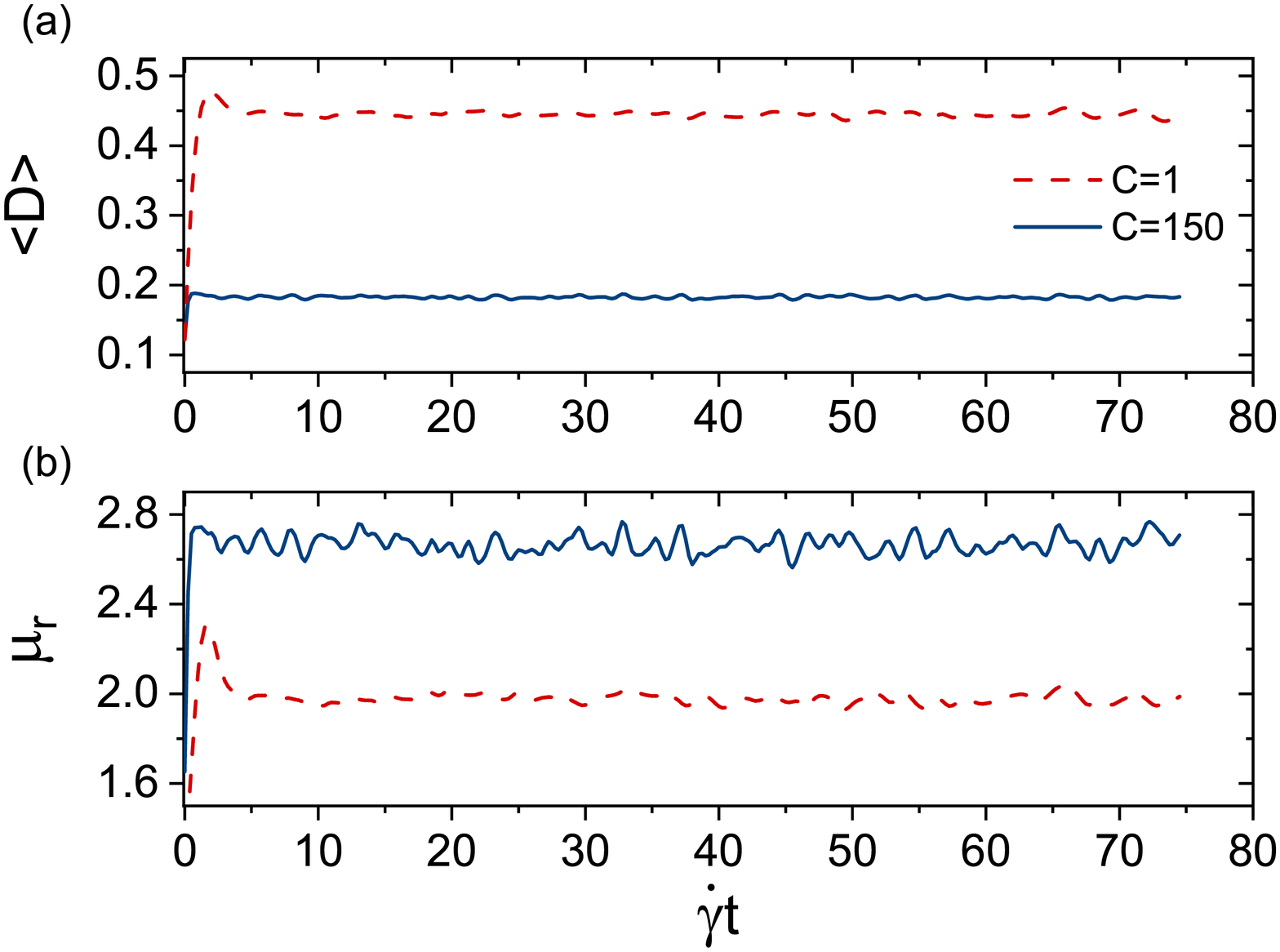}
        \caption{History of the mean deformation (a) and the relative viscosity (b) of the capsules for a semi-dilute suspension ($\phi=0.3$) and two different membrane inextensibilities $(C=1,150)$.
        }
        \label{fig:fig11}
\end{figure}

Figure \ref{fig:fig11} depicts the time evolution of the mean deformation and relative viscosity of a suspension of capsules in the semi-dilute limit with $\phi=0.3$ for two different values of the membrane inextensibility $(C=1,150)$, and a fixed capillary number $(Ca=0.5)$. The time average is performed on the interval where the Taylor deformation index averaged over the number of particles reaches a steady-state value with fluctuations less than $1\%$. In term of strain units, we have observed that the transient time spans over the first $5$ to $7 \dot{\gamma}t$.

\begin{table}
  \begin{center}
      \begin{ruledtabular}
          \begin{tabular}{lccccc}
              $N_f$               & $320$ & $1280$ & $5120$  \\[3pt]
              $r_1/r$             & $1.429 \pm 0.027$  & $1.445 \pm 0.028$  & $1.442 \pm 0.027$\\
              $r_2/r$             & $1.095 \pm 0.012$  & $1.117 \pm 0.0145$  & $1.116 \pm 0.0138$\\
              $r_3/r$             & $0.61 \pm 0.023$  & $0.612 \pm 0.024$  & $0.619 \pm 0.024$\\      
              $D$                 & $0.401 \pm 0.024$  &  $0.404 \pm 0.025$ & $ 0.398 \pm 0.024$ \\      
              $\theta/\pi$        & $0.138 \pm 0.007$  & $0.139 \pm 0.007$  &$0.139 \pm 0.007$ \\      
              $[\mu]$             & $2.109 \pm 0.124$  & $2.226 \pm 0.129$  & $2.25 \pm 0.133$\\
          \end{tabular}
      \end{ruledtabular}
  \caption{Steady-state shape and rheology quantities (time averaged) for $Ca=0.5$ and $C=1$.}
  \label{tab:mesh_quality}
  \end{center}
\end{table}

\subsection{\label{sec:appendix-grid}Grid independency}
In order to check that the system size in the periodic directions is large enough 
to ensure grid independency of the results, we test here the behaviour of two relevant observables,
namely the mean Taylor deformation parameter, $\langle D \rangle$, and the relative viscosity,
$\mu_r$.
We performed simulations at changing the box size from $8r \times 8r \times 16r$ to 
$64r \times 64r \times 16r$, for fixed $C=150$, $Ca=0.5$, 
$\phi=0.3$ and wall-to-wall distance $L=16r$. We have varied the number of particles ($N_p$) to keep a fixed volume fraction. 
For the sake of comparison, we look at the relative (percentual) errors for 
for the mean Taylor deformation parameter and the relative viscosity, defined as
\begin{equation}
    \epsilon_D = \frac{\langle D \rangle- \langle D \rangle^{(0)}}{\langle D\rangle^{(0)}}, 
\end{equation}
and
\begin{equation}
    \epsilon_{\mu_r} = \frac{\mu_r - \mu_r^{(0)}}{\mu_r^{(0)}},
\end{equation}
respectively ($\langle D \rangle^{(0)}$ and $\mu_r^{(0)}$ being the values corresponding to the cubic reference case, used throughout the paper).
The results are reported in Table \ref{tab:rel_error_sys_size}. 
All relative errors are small and decrease from 
$0.47\%$ to $0.38\%$ (for the deformation parameter) and 
from from $3\%$ to $2\%$ (for the relative viscosity), denoting a satisfactory degree of 
convergence for the resolution employed.
\begin{table}
    \begin{center}
        \begin{ruledtabular}
            \begin{tabular}{lccccc}
              Box size  & $N_p$ & $\epsilon_{D}$   &   $\epsilon_{\mu_r}$  \\[3pt]
              \hline
               $8r\times8r\times16r$    & 73   & 0.47\% & 3\% \\
               $32r\times32r\times16r$  & 1174 & 0.62\% & 2.3\% \\
               $64r\times64r\times16r$  & 4696 & 0.38\% & 2\% \\
            \end{tabular}
        \end{ruledtabular}
  \caption{Relative errors on $\langle D \rangle$ and $\mu_r$ for different system sizes}
  \label{tab:rel_error_sys_size}
  \end{center}
\end{table}
\subsection{\label{sec:appendix-repulsive_force}Effect of the short range repulsive force}
\begin{figure}
        \centering
        \includegraphics[width = 0.65\textwidth]{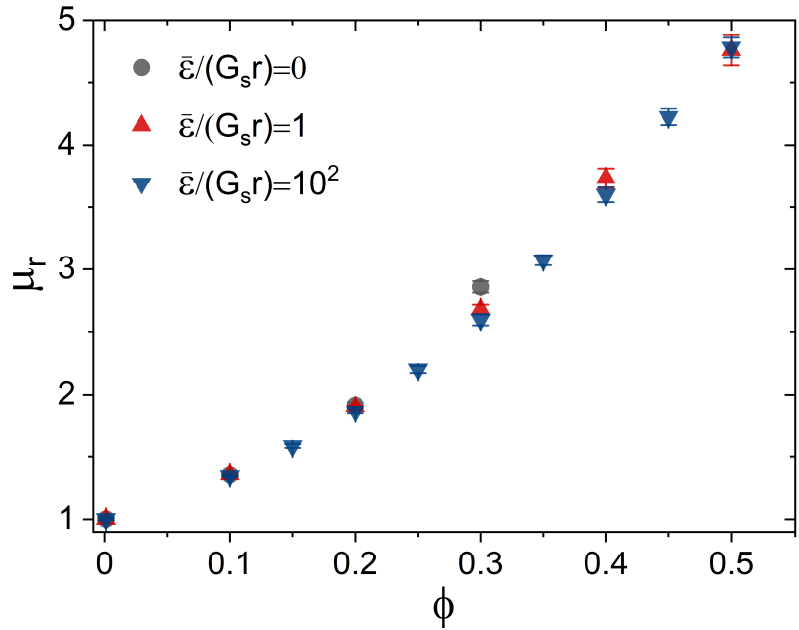}
        \caption{Effect of the short range repulsive force on the relative viscosity of a suspension of capsules. 
        }
        \label{fig:fig12}
\end{figure}

A well-known limitation of the LB-IBM scheme is the interpolation of the membrane velocity from the surrounding fluid when the distance between two boundary nodes is below the lattice resolution limit which can result into a permanent overlap between neighboring nodes \citep{kruger2012computer}. The probability of such event to occur increases with the volume fraction of the suspension. To prevent such situation a short range repulsive force must be applied on neighboring nodes from different membranes when the node-to-node distance ($d_{ij}$) is below one lattice spacing ($\Delta x$), whereas it vanishes for $d_{ij} > \Delta x$ (see equation (\ref{eq:frep})). 
Without any short-range repulsion, we do indeed observe crossing particles for $\phi>0.3$. Such force is, therefore, needed, however its details do not
affect the macroscopic behaviour of the suspension, within the range of volume fractions explored in this study. 
To show this, we report in figure \ref{fig:fig12} results on the relative viscosity from simulations 
with $C=150$, $Ca=0.5$ and $\phi \in [10^{-3}, 0.5]$, for three different values of the force amplitude: $\bar{\epsilon}/(G_s r) = 0$ (absence 
of force), $\bar{\epsilon}/(G_s r) = 1$ and $\bar{\epsilon}/(G_s r) = 10^2$ (the value used in all simulations throughout the paper). 
Only a slight deviation (of less than $10\%$) can be detected for $\bar{\epsilon}/(G_s r) = 0$ at $\phi=0.3$ 
(for larger $\phi$ data are not available because of the occurrence of crossings), otherwise all sets of data basically overlap, within error bars.


\end{document}